\documentclass[12pt,preprint]{aastex}

\usepackage{subfigure}
\usepackage{color}

\usepackage{amsmath,amssymb,amsfonts}
\usepackage{times}
\def\vector#1{\mbox{\boldmath $#1$}}  

\shorttitle{}
\shortauthors{Kawabata et al.}

\begin{document}


\title{Non-potential field formation in the X-shaped quadrupole magnetic field configuration}


\author{Y. Kawabata\altaffilmark{1, 2}, S. Inoue \altaffilmark{3, 4}, and T. Shimizu \altaffilmark{2, 1}}
\affil{Department of Earth and Planetary Science, The University of Tokyo}

\email{kawabata.yusuke@ac.jaxa.jp}


\altaffiltext{1}{Department of Earth and Planetary Science, The University of Tokyo, 7-3-1 Hongo, Bunkyo-ku, Tokyo 113-0033, Japan}
\altaffiltext{2}{Institute of Space and Astronautical Science, Japan Aerospace Exploration Agency, 3-1-1 Yoshinodai, Chuo, Sagamihara, Kanagawa 252-5210, Japan}
\altaffiltext{3}{Max-Planck-Institute for Solar System Research, Justus-von-Liebig-Weg 3 D-37077 G$\rm \ddot{o}$ttingen Germany}
\altaffiltext{4}{Institute for Space-Earth Environmental Research, Nagoya University, Furo-Cho, Chikusa-ku Nagoya 464-8601, Japan}
\begin{abstract}
Some types of solar flares are observed in X-shaped quadrupolar field configuration. To understand the magnetic energy storage in such a region, we studied non-potential field formation in an X-shaped quadrupolar field region formed in the active region NOAA 11967, which produced three X-shaped M-class flares on February 2, 2014. Nonlinear force-free field modeling was applied to a time series of vector magnetic field maps from the Solar Optical Telescope on board {\it Hinode} and Helioseismic and Magnetic Imager on board {\it Solar Dynamics Observatory}. 
Our analysis of the temporal three-dimensional magnetic field evolution shows that the sufficient free energy had already been stored more than 10 hours before the occurrence of the first M-class flare and that  the storage was observed in a localized region. In this localized region, quasi-separatrix layers (QSLs) started to develop gradually from 9 hours before the first M-class flare.  One of the flare ribbons that appeared in the first M-class flare was co-spatial with the location of the QSLs, suggesting that the formation of the QSLs is important in the process of energy release. These QSLs do not appear in the potential field calculation, indicating that they were created by the non-potential field. The formation of the QSLs was associated with the transverse photospheric motion of the pre-emerged flux and the emergence of a new flux. This observation indicates that the occurrence of the flares requires the formation of QSLs in the non-potential field in which free magnetic energy is stored in advance.
\end{abstract}






\section{Introduction}
Solar flares are explosive energy release events in the solar atmosphere. The released energy is thought to be the magnetic energy stored in the upper atmosphere. The emergence of twisted magnetic flux \citep{1996ApJ...462..547L,1998ApJ...492..804E} and transverse photospheric motions \citep{1982SoPh...79...59K, 2014A&A...562A.110L} have been considered as the key physical processes for storing the magnetic energy.
Magnetic reconnection is thought to play an important role in converting magnetic energy to kinetic and thermal energy in the solar atmosphere and previous observations have provided some pieces of evidence on magnetic reconnection during solar flares \citep{1992PASJ...44L..63T,1994Natur.371..495M,1996ApJ...456..840T}. Solar flares are often accompanied by coronal mass ejections (CME) which give influence to the Earth's magentosphere and lead to geomagnetic storms. The analysis of the three-dimensional (3D) magnetic field structure development in the corona is essential for understanding energy storage and magnetic reconnection.

In the solar corona, the magnetic pressure is dominant; thus the plasma  $\beta (=8\pi  p/B^2)$, which is the ratio between the gas pressure $p$ and the magnetic pressure $B^2/8\pi$, where $B$ is the magnetic flux density, is sufficiently small \citep{2001SoPh..203...71G}. 
In such circumstances, the gas pressure can be neglected and the equilibrium is achieved when the Lorentz force vanishes, i.e., the magnetic tension and the magnetic pressure are balanced. That is,
\begin{equation}
\vector{j}\times\vector{B}=0,
\label{fff}
\end{equation}
where $\vector{j}$ is the current density, derived from $\nabla \times \vector{B}$. To reproduce the magnetic fields in active regions, we must numerically solve the nonlinear equation (\ref{fff}) based on the magnetic field at the photosphere.
The lowest energy state of the magnetic field configuration is called  the potential (current-free) field.
When the magnetic field deviates from the potential field, it has the excess energy, which is called free energy and can be used for energy release in solar flares \citep{2012LRSP....9....5W}. 
To estimate the free energy quantitatively, we must obtain the 3D coronal magnetic field. Although we can derive the photospheric magnetic field from observing Stokes vector (i.e., the polarimetric signals generated by Zeeman effect), the measurement of the magnetic field in the chromosphere and the corona is challenging because of the low signal of polarization \citep{2015SSRv..tmp..115L}. 
Thus, nonlinear force-free field (NLFFF) modeling is currently used as a powerful tool for obtaining insights into the magnetic structure in the corona, in addition to morphological information of the coronal magnetic structures, which is acquired from extreme ultraviolet (EUV) and soft X-ray imaging observations. 
The main concept of NLFFF modeling is the extrapolation of the coronal magnetic field from the spatial map of the magnetic field at the photosphere \citep{2012LRSP....9....5W}. 

 Previous studies of NLFFF modeling have mainly been conducted in sheared bipolar regions \citep{2008ApJ...675.1637S,2011ApJ...738..161I,2016A&A...591A.141J}. Some recent studies focused on the reconnection process in the quadrupolar region \citep{2012ApJ...760..101W,2016ApJ...825..123J,2016ApJ...829L...1L}.  
 However, the mechanism and location of free energy storage and formation of the non-potential magnetic field in the quadrupolar region remain unclear. The main purpose of this study is to investigate the energy storage process in an X-shaped quadrupolar magnetic field region by applying our NLFFF modeling to the active region NOAA 11967, which produced three M-class flares in an X-shaped quadrupolar magnetic configuration formed in the active region. This X-shaped quadrupolar region was recently studied by \cite{2016arXiv160902713L}, which investigated the temporal change of null point height based on the potential field modeling. 
 They applied a NLFFF modeling to only one magnetogram, which was acquired just before the occurrence of flares, to confirm that a current layer was formed at the intersection of two quasi-separatrix layers (QSLs). In our study, we apply our NLFFF modeling to a time series of photospheric magnetic field data and  explain how the non-potential coronal structure was formed. 

Before investigating the temporal evolution of the 3D magnetic field, we evaluated the validity of the force-free modeling. \cite{2009ApJ...696.1780D} modeled the coronal magnetic field by applying a suite of NLFFF algorithms to the photospheric vector field. They validated the results against images of coronal loops observed in EUV or X-ray images. In this study, we applied a different method in order to validate the accuracy of our NLFFF modeling: we compared the derived magnetic field connectivity with the location of flare ribbons.

This paper is organized as follows: The observations and data analysis are described in Section 2, and the method of deriving the NLFFF and QSLs is explained in Section 3. Section 4 presents the results, followed by the discussion in Section 5 and the conclusions in Section 6.


\section{Observations and  Data Analysis}
\begin{figure}
\centering
 \includegraphics[width=0.75\columnwidth,clip,bb=0 0 472 754]{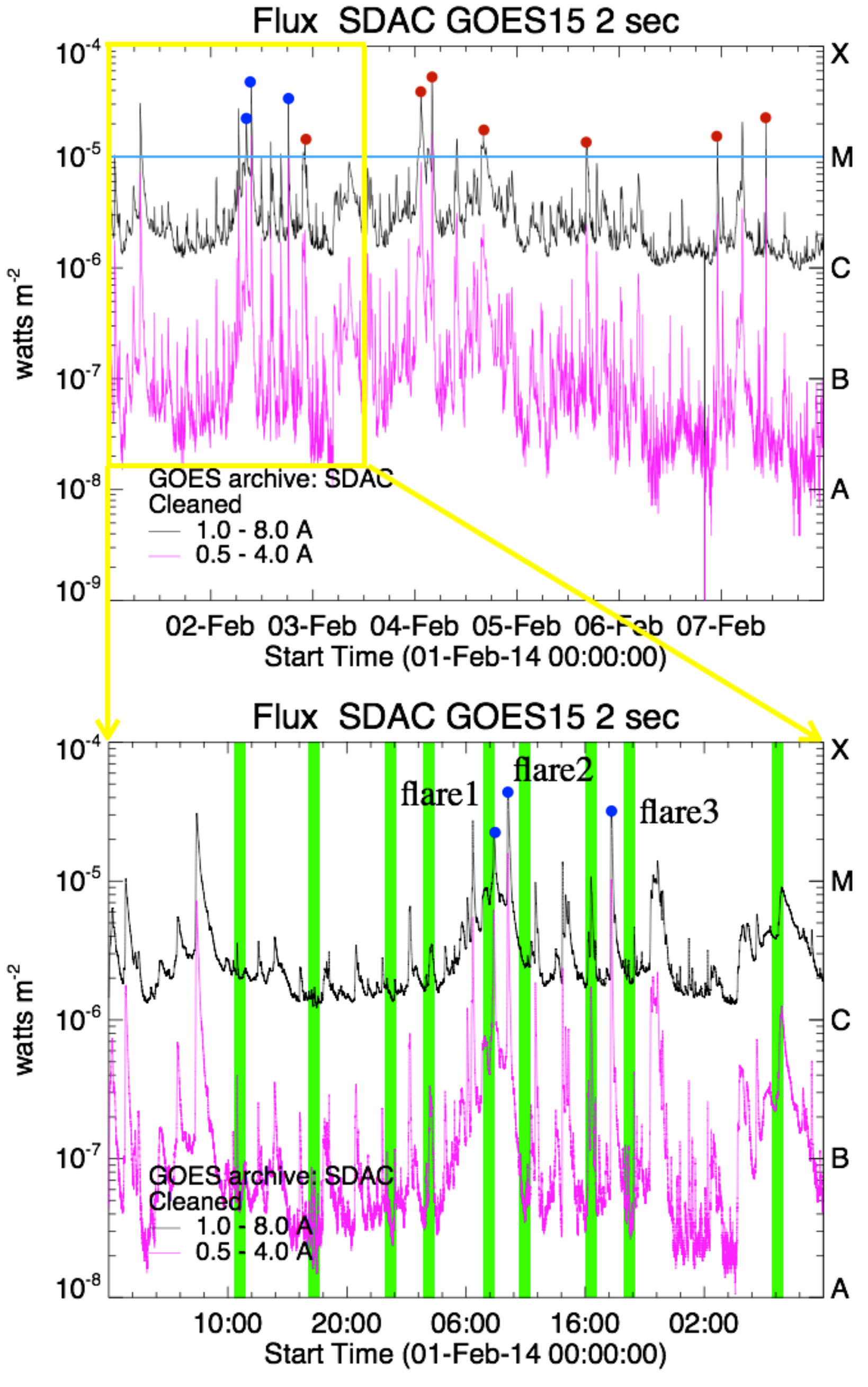}
 \caption{ The black line shows the 1-8 \AA \ flux and the purple line shows the 0.5-4 \AA \ flux obtained by GOES. NOAA 11967 produced 10 M-classes flares from February 1, 2014 to February 8, 2014, which are shown by the red and blue circles. M-class is defined above the energy flux given by the blue line. The green lines show the timing of the observed {\it Hinode} SOT/SP data and the blue circles correspond to the flare timing in Figures \ref{obs1}, \ref{obs2}, and \ref{obs3}.}
 \label{goes}
\end{figure}

\begin{figure}
\centering
 \includegraphics[width=\columnwidth,clip,bb=0 0 1123 860]{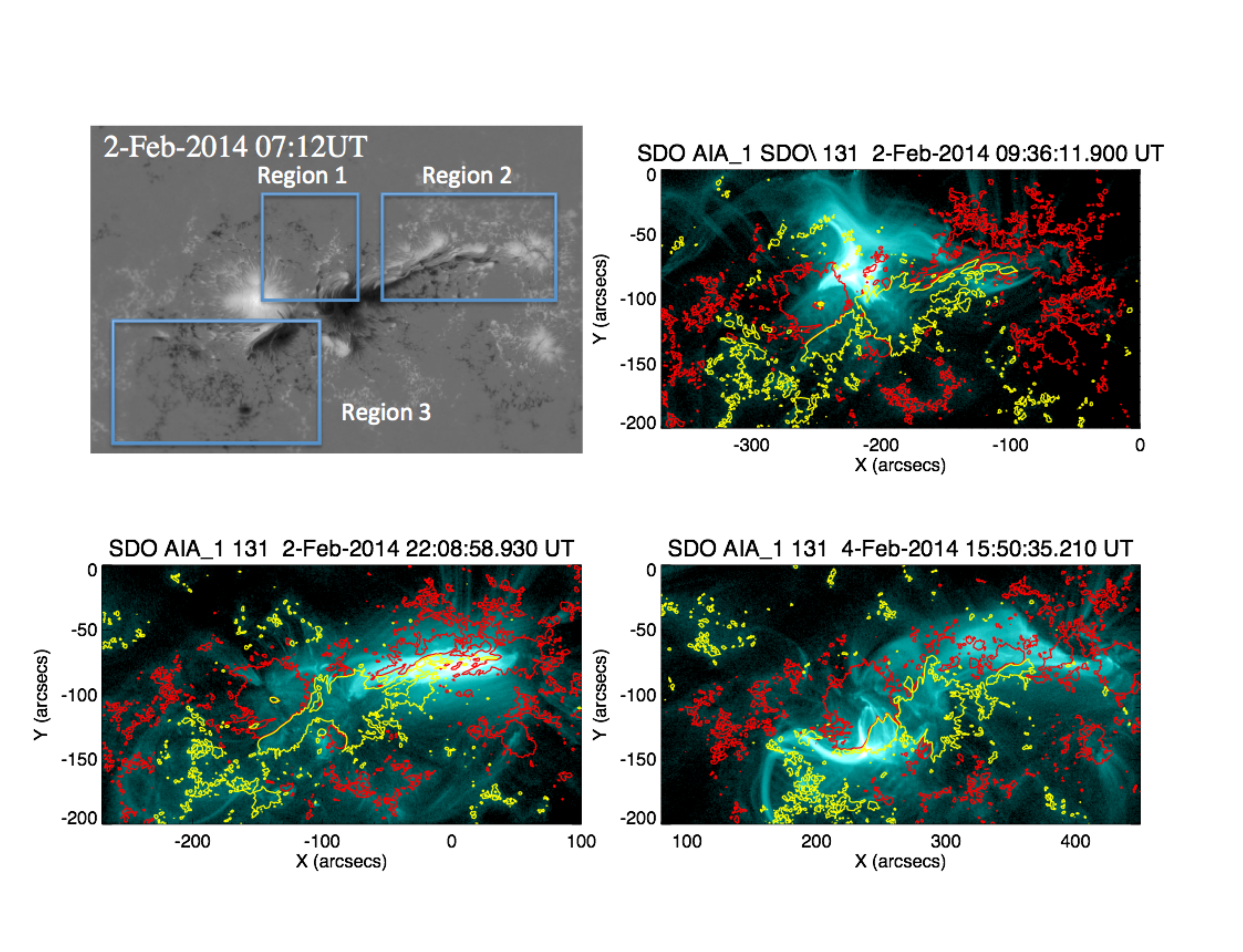}
 \caption{Upper left: the line-of-sight magnetic flux density of NOAA 11967 obtained by the HMI. Upper right: the EUV image during a flare in region 1 obtained by AIA 131 \AA. The color contours show the line-of sight magnetic flux density obtained by HMI (red: 300 G, yellow: -300 G) at almost same timing as the EUV observation. Bottom left: the EUV image during a flare in region 2 obtained by AIA 131 \AA. Bottom right: the EUV image during a flare in region 3 obtained by AIA 131 \AA.}
 \label{flares}
\end{figure}

The active region NOAA 11967 appeared at the east limb of the Sun at the end of January 2014 and developed its magnetic structure during its passage through the disk. Figure \ref{goes} shows the soft X-ray fluxes from February 1, 2014 to February 8, 2014, measured with {\it Geostationary Environmental Satellite} ({\it GOES}).
Between February 1, 2014 and February 8, 2014, NOAA 11967 produced 10 M-class flares, as shown by the red and blue circles in the upper panel of Figure \ref{goes}. It must be noted that the spikes without red and blue circles in Figure \ref{goes} were produced at other active regions. The bottom panel of Figure \ref{goes} shows the soft X-ray fluxes from February 1 to February 3, which corresponds to the time range indicated by the yellow rectangle in the upper panel. The green lines show the timing of the observation with Solar Optical Telescope/Spectropolarimeter (SOT/SP) on board  {\it Hinode} and the blue circles correspond to the flare timing in Figures \ref{obs1}, \ref{obs2}, and \ref{obs3}. As shown in Figure \ref{flares} and Table \ref{events}, three regions (region 1, 2, and 3) in the active region produced M-calss flares. The upper left panel of Figure \ref{flares} shows the line-of-sight magnetic field obtained by the Helioseismic and Magnetic Imager \citep[HMI;][]{2012SoPh..275..229S} on board {\it Solar Dynamics Observatory (SDO)}. The upper right panel (region1) and lower panels (left:region2, right:region3) show Extreme ultraviolet (EUV) images (131 \AA) observed with the Atmospheric Imaging Assembly \citep[AIA;][] {2012SoPh..275...17L} on board {\it SDO}.
In contrast to regions 2 and 3, the photospheric magnetic field in the region 1 clearly shows a multipole magnetic field.
In this study, we focus on the region 1, which produced three M-class flares beginning at 08:03 (M2.2, S10E14), 09:24 (M4.4, S11E13), and 18:05UT (M3.1, S10E08) on February 2, 2014. 

\begin{table}
\centering
\begin{tabular}{ccc} \hline \hline
date & X-ray class & Region \\
\hline
2014/02/02 08:20 & M2.2 & R1 \\
2014/02/02 09:31 & M4.4 & R1 \\
2014/02/02 18:11 & M3.1 & R1 \\
2014/02/02 22:04 & M1.3 & R2 \\
2014/02/04 04:00 & M5.2 & R3 \\
2014/02/04 09:49 & M1.4 & R3 \\
2014/02/04 15:30 & M1.5 & R2 and R3 \\
2014/02/05 16:20 & M1.3 & R2 \\
2014/02/06 23:05 & M1.5 & R3 \\
2014/02/07 04:56 & M2.0 & R3 \\ \hline\hline
\end{tabular}
\caption{M class flares produced by NOAA 11967}
\label{events}
\end{table}

\subsection{{\it Hinode} SOT Observations}

The SOT \citep{2008SoPh..249..167T,2008SoPh..249..221S,2008SoPh..249..197S,2008SoPh..249..233I} on board {\it Hinode} has two focal plane instruments, i.e., the Filtergraph (FG) and the SP \citep{2013SoPh..283..579L}. In this study, we used series a of the SP data.

 The SP performs spectropolarimetric observations with  two magnetically sensitive Fe I lines at 6301.5 \AA \ and 6302.5 \AA. We used nine vector magnetic field maps derived from the SP observations. The maps have an effective pixel size of 0$\arcsec$.3 with a field of view (FOV) of 280$\arcsec \times$130$\arcsec$. 
The spectral sampling was 21.549 m\AA \ pixel$^{-1}$.
The SP maps were taken with fast-mapping mode by slit scanning over almost the entire active region every two or three hours. Because we want to investigate the difference between the global coronal 3D magnetic field during flare productive and  less productive periods, we used nine SP maps obtained at, 10:42-11:35 UT, 16:54-17:45 UT, and 23:19-00:14 UT on February 1, 2014, and  02:20-03:15 UT, 07:12-08:07 UT, 10:20-11:15 UT, 16:00-16:52 UT, and 19:20-20:15 UT on February 2, 2014, and 07:50-08:45 UT on February 3. As shown in Figure \ref{goes}, the observation period was between 21 hours before flare 1 and 14 hours after flare 3.

For the calibration of the Stokes profiles, we used the Solarsoft routine SP$\_$PREP \citep{2013SoPh..283..601L}.
When we derived physical parameters from the Stokes profiles, we solved the radiative transfer equation \citep{2007insp.book.....D},
\begin{equation}
\frac{d\vector{I}}{d\tau}=\vector{K}(\vector{I}-\vector{S}),
\end{equation} 
where $\vector{I} \equiv (I,Q,U,V)$ is the Stokes vector, $\tau$ is the optical depth, $\vector{K}$ is the propagation matrix, and $\vector{S}$ is the source function.
We applied  the {\it Milne-Eddington atmosphere} (ME) model to the two lines (Fe I doublet at 6301.5\AA \ and 6302.5 \AA) to derive the physical parameters by a nonlinear least square fitting using the code based on MELANIE \citep{2001ASPC..236..487S}. This simultaneous inversion has been shown to provide better results than the use of  only a single line \citep{1994SoPh..155....1L}. 
The 180 degree ambiguity in the transverse magnetic field direction was solved with the minimum energy ambiguity resolution method \citep{1994SoPh..155..235M,2009ASPC..415..365L}.

\subsection{{\it SDO} Observations}

The AIA was used to investigate the coronal and chromospheric structures of the flaring region in NOAA 11967. In this study, we use 1600 \AA \ ,which contains C IV line and continuum, for investigating the behaviors of the flare ribbons in the chromosphere and 131 \AA \ , which contains Fe VII, XX, and XXIII lines, for the coronal hot plasma in the flaring region. The pixel size was 0\arcsec.6 and the temporal cadence was 12s for 131 \AA \ images and 24s for 1600 \AA \ images. The HMI on board {\it SDO} provided the full-disk photospheric magnetic field.
The HMI measures polarimetric signals ($I \pm Q, I \pm U, I \pm V$) of magnetic sensitive Fe I 6173 \AA \ line at 6 narrow bands (band width 76 m\AA \ +/- 10 m\AA \ ) in the line in a 135-second sequence and derive the magnetic field vectors, which are averaged over 12 minutes to increase the signal-to-noise ratio. 

The Solarsoft routine AIA$\_$PREP was used for the AIA data calibration and the alignment between the AIA and HMI data. We investigated the coronal and chromospheric  dynamics  by visually inspecting of the EUV(131 \AA) and UV (1600 \AA) data acquired near the occurrence of flare 1 (08:00UT-09:00UT), flare 2 (09:00UT-10:00UT), and flare 3 (18:00UT-18:30UT).

\subsection{Bottom Boundary for NLFFF modeling}
\begin{figure}
\centering
\subfigure[]{
 \includegraphics[width=0.72\columnwidth,clip,bb=0 0 1417 1133]{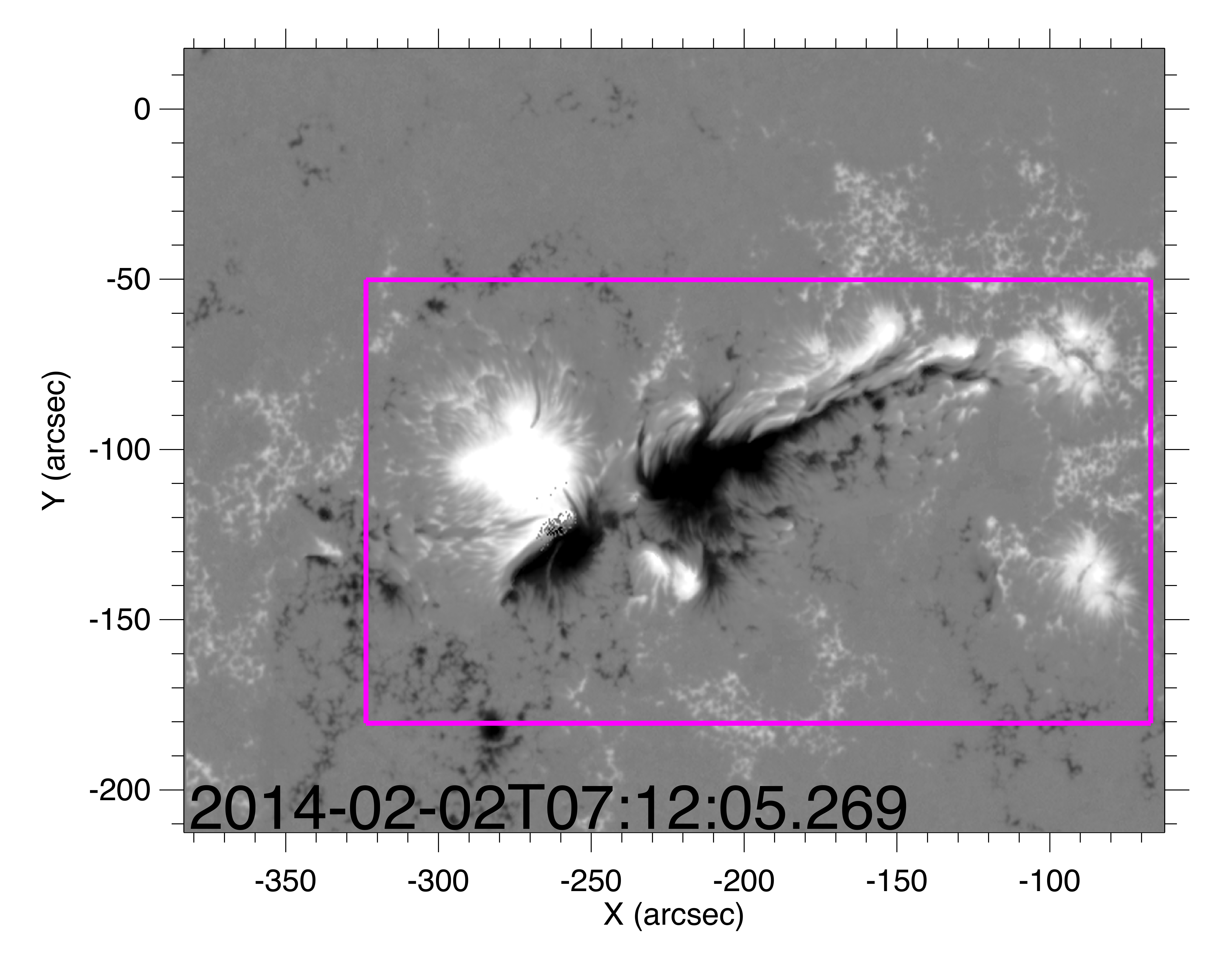}}
  \subfigure[]{
 \includegraphics[width=0.72\columnwidth,clip,bb=0 0 1417 1133]{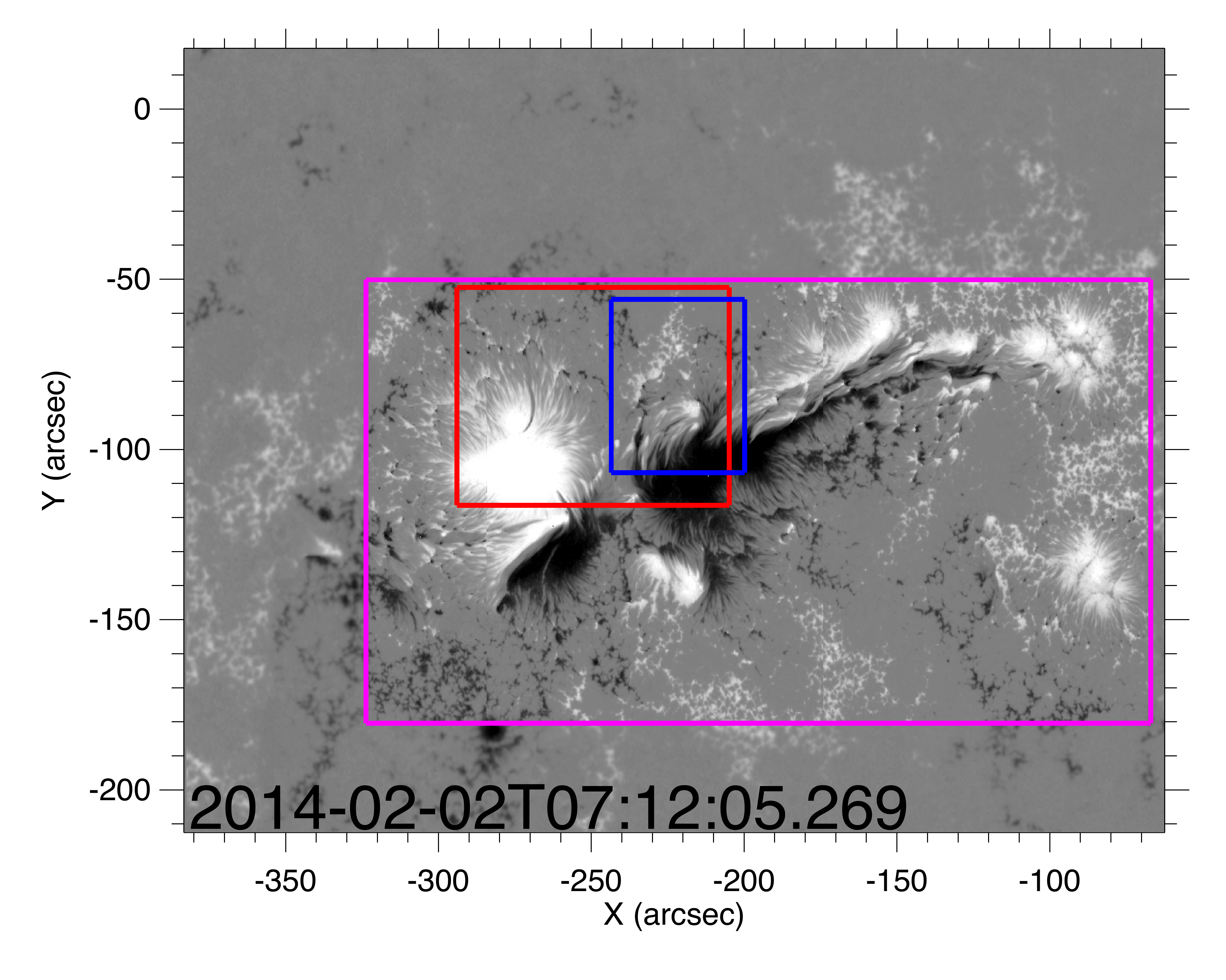}}
\caption{(a) The vertical magnetic flux density obtained by {\it SDO}/HMI, (b) Vertical magnetic flux density obtained by {\it Hinode}/SOT after increasing the FOV with  {\it SDO}/HMI. The map in (b) was used in the nonlinear force-free field modeling. The magenta square shows the FOV of {\it Hinode}/SOT SP. The red, and  blue squares in Figure \ref{boundary} (b) show the FOV in Figures \ref{fen_evo} and \ref{qsl_evolution}, and Figures \ref{photo_evo} and \ref{trans_evo}, respectively.
}
\label{boundary}
\end{figure}

For the HMI data, there is a data product, called Space-weather HMI Active Region Patches (SHARP), which  provides vector magnetic field data on a 12 minutes cadence and automatically identifies and tracks active regions \citep{2014SoPh..289.3549B}.
We used these vector field data to increase the narrow FOV of Hinode/SP when extrapolating the coronal magnetic field by NLFFF modeling (Figure \ref{boundary}). 
When we created the boundary data for the modeling, we converted the magnetic field vector map in the line-of-sight coordinates to that in the local frame coordinates and resized the HMI pixel size to the SP pixel size with a cubic polynomial interpolation using 16 neighboring points.
The resized HMI data were co-aligned to the SP data by applying a cross-correlation technique.
One of the co-aligned HMI data is shown in Figure \ref{boundary} (a).
To obtain wider FOV data with higher polarimetric accuracy, the co-aligned HMI data were added to around the SP map. 
The 2$\times$2 binned data of the combined SP/HMI maps, one of which is shown in Figure \ref{boundary} (b), were used as the boundary condition for the nonlinear force-free field modeling. The pixel size is 0\arcsec.6 $\times$ 0\arcsec.6. The magenta, red, and  blue squares in Figure \ref{boundary} (b) shows the FOV of the {\it Hinode}/SOT SP, Figures \ref{fen_evo} and \ref{qsl_evolution}, and Figures \ref{photo_evo} and \ref{trans_evo}, respectively.

\section{NLFFF and Squashing Factor}

 In this section, we describe the numerical methods applied in NLFFF modeling \citep{2014ApJ...780..101I} and the squashing factor $Q$ \citep{1999ESASP.448..715T}, which will be used in the following sections.

\subsection{Force-free field modeling \label{fffm}}
Several methods are used to solve the boundary value problem and obtain a force-free field \citep{2012LRSP....9....5W}; here, we used the magnetohydrodynamics (MHD) relaxation method. This method was first developed by\cite{1994ASPC...68..225M} and used zero-$\beta$ time-dependent MHD codes to achieve stationary equilibrium. Their calculation begins with a potential field calculated from the normal components of the magnetic field at the photosphere, and they control the transverse electric field while maintaining $B_{z}$ stable;  eventually, they obtain a force-free state. 

We used the code developed by \cite{2014ApJ...780..101I} and also see in \cite{2016PEPS....3...19I}, which improved \cite{1994ASPC...68..225M}. To achieve a force-free state, we solved the following zero-$\beta$ MHD equations in the Cartesian coordinates,
\begin{eqnarray}
\frac{\partial \vector{v}}{\partial t}&=&-(\vector{v}\cdot \vector{\nabla})\vector{v}+\frac{1}{\rho}\vector{j}\times\vector{B}+\nu\nabla^2\vector{v}, \label{momentum} \\
\frac{\partial \vector{B}}{\partial t}&=&\nabla \times (\vector{v}\times \vector{B}-\eta \vector{j})-\nabla \phi, \label{induction}\\
\vector{j}&=& \nabla \times \vector{B}, \label{ampere}\\
\frac{\partial \phi}{\partial t} &+&c_{h}^2\nabla\cdot \vector{B}=-\frac{c_{h}^2}{c_{p}^2}\phi, \label{9wave}
\end{eqnarray}
where $\rho$ is the pseudo density, which is assumed to be proportional to $|\vector{B}|$,  $\phi$ is the convenient potential for $\nabla \cdot \vector{B}$ cleaning and $\nu$ is the viscosity, which was set to be a constant ($1.0 \times 10^{-3}$). The length, magnetic field, velocity, and time were normalized by $L_{0}=157 $ Mm and $B_{0}=4000$ G, $V_{\rm A} \equiv B_0/(4\pi \rho_0)^{1/2}$, and $\tau_{A} \equiv L_{0}/V_{A}$, where $V_{A}$ is the Alfv$\rm {\acute{e}}$n velocity. Eqns. (\ref{momentum}), (\ref{induction}), (\ref{ampere}) and  (\ref{9wave}) are the equation of motion,  the induction equation, the ${\rm Amp\grave{e}re's}$ law, and $\nabla\cdot\vector{B}$ cleaning scheme introduced by \cite{2002JCoPh.175..645D}, respectively. The parameters $c_{p}^2$ and $c_{h}^2$ are the advection and diffusion coefficients, respectively, which were fixed at $0.1$ and $0.04$, correspondingly. The non-dimensional resistivity $\eta$ is given by
\begin{equation}
\eta=\eta_{0}+\eta_{1}\frac{|\vector{j}\times\vector{B}||\vector{v}|^2}{|\vector{B}|^2},
\end{equation}
where $\eta_{0}$ and $\eta_{1}$ are fixed at $5.0\times 10^{-5}$ and $1.0\times10^{-3}$ in non-dimensional units.  

The velocity field was adjusted as below to avoid increasing when the value of $v^{*}$ becomes larger than the value of $v_{\rm{max}}$,
\begin{equation}
\vector{v} \rightarrow \frac{v_{{\rm max}}}{v^{*}}\vector{v},
\end{equation}
where $v^{*}=|\vector{v}|/|\vector{v_{A}}|$, where $v_{{\rm max}}=0.01$. 

The initial condition was the potential field calculated from the normal component of the observed magnetic field  and the calculation used Green's function method \citep{1964NASSP..50..107S,1982SoPh...76..301S}.
The magnetic field at the top and side boundaries was maintained at the initial state (or potential field) and the normal component of the magnetic field at the bottom boundary was also fixed. 
We varied the transverse component at the bottom boundary $\vector{B}_{BC}$ as follows,
\begin{equation}
\vector{B}_{BC}=\gamma\vector{B}_{obs}+(1-\gamma)\vector{B}_{pot},
\label{bottombc}
\end{equation}
where $\vector{B}_{obs}$ and $\vector{B}_{pot}$ are the transverse component of the observational and potential fields, respectively. 
We increased $\gamma=\gamma+d\gamma$ when $\int |\vector{j}\times\vector{B}|^2dV$ was reduced below a critical value. In this study we set $d\gamma=0.1$ and when $\gamma$ became equal to $1$, $\vector{B}_{BC}$ was consistent with the observed field.

Spatial derivatives were calculated by integrating the second-order central differences and  temporal derivatives were integrated by the Runge-Kutta-Gill method to fourth-order accuracy. 
We used the combined SP and HMI data at the boundary condition on the bottom, which is shown in Figure \ref{boundary} (b). 
The numerical domain was set to $(0,0,0) <(x,y,z)<(1.5,1.0,0.5)$, which was resolved by $540\times360\times180$ nodes. The relaxation for each map was computed using 144 processors  of the JAXA Supercomputer System generation 2 (JSS2) and the computation time was approximately 5 hours.
To visualize the field lines, we used the UCAR's Vapor 3D visualization package (www.vapor.ucar.edu).

\subsection{Squashing factor \label{analysis:sq}}
In this study, we attempted to detect the formation of a flare-productive 3D magnetic field structure, i.e., the magnetic field configuration, in which the magnetic reconnection is likely to occur. A necessary and sufficient condition for reconnection in 3D is the existence of a thin current layer,  in which the ideal MHD breaks down. \cite{1995JGR...10023443P} proposed a model of reconnection in the absence of null points. 
They suggested that the current layer is formed in quasi-separatrix layers (QSLs), in which the gradient of field line linkage is steep but continuous. To define QSLs, the norm $N$ has been proposed \citep{1995JGR...10023443P,1996A&A...308..643D}, which is,
\begin{equation}
N=\sqrt{\left[\left(\frac{\partial X}{\partial x}\right)^2+\left(\frac{\partial X}{\partial y} \right)^2+\left(\frac{\partial Y}{\partial x}\right)^2+\left(\frac{\partial Y}{\partial y} \right)^2\right]},
\end{equation}
where $(X,Y)$ is another footpoint of the field line connecting from a selected footpoint at (x,y). 
$N$ is the norm of the displacement gradient tensor and is evaluated only at the boundary. 
They claimed that QSLs exist at the locations of high $N$. 
$N$ is non-dimensional and independent of the coordinate system. However, the norm has a problem in defining QSLs. 
If we calculate the norm at different footpoints of the same field line, the values are generally different, making the determination of the QSLs ambiguous. 
To solve this problem, a new definition of QSLs called squashing factor $Q$ was proposed by  \cite{1999ESASP.448..715T} and improved in later studies \citep{2002JGRA..107.1164T,2007ApJ...660..863T,2012A&A...541A..78P}. 
The squashing factor $Q$ is defined as

\begin{equation}
Q=\frac{N^2}{\Delta},
\label{eq:squash}
\end{equation}
where the Jacobian matrix $D$ of the field-line mapping and its determinant $\Delta$ are
\begin{equation}
D=\left(
\begin{array}{cc}
\partial X/\partial x&\partial X/\partial y \\
\partial Y/\partial x&\partial Y/\partial y
\end{array}
\right)=\left(
\begin{array}{cc}
a&b\\
c&d
\end{array}
\right),
\end{equation}
\begin{equation}
\Delta=ad-bc.
\end{equation}
The squashing factor $Q$ is defined by the product of the norms $N_{+}$ and $N_{-}$, which are calculated from each footpoints of the same field line. There is a relation between $N_{+}$ and $N_{-}$,
\begin{equation}
N_{+}=N, \ \ \ N_{-}=\frac{N}{\Delta}.
\end{equation}
Therefore, $Q$ values calculated from the footpoints of the same field line are equal. \cite{2002JGRA..107.1164T} showed that the determinant of Jacobian matrix ($\Delta$) is related to the ratio of the normal component of the magnetic field at the boundary,
\begin{equation}
|\Delta|=\frac{|B_{z,+}|}{|B_{z,-}|}.
\label{eq:delta}
\end{equation}
Although the squashing factor $Q$ is originally defined only at the boundary, \cite{2012A&A...541A..78P} extended it to a 3D domain assuming that the squashing factor is invariant along a field line. 

By using eqns. (\ref{eq:squash}) and (\ref{eq:delta}), we calculated the squashing factor in a 3D domain using the results of the force-free modeling. 
We integrated the field lines using the fourth-order Runge-Kutta method and chose the step size by adaptive step-size control.


\section{Results}
\subsection{Properties of the flares from observations}
\begin{figure}
\centering
 \includegraphics[width=0.9\columnwidth,clip,bb=0 0 1034 793]{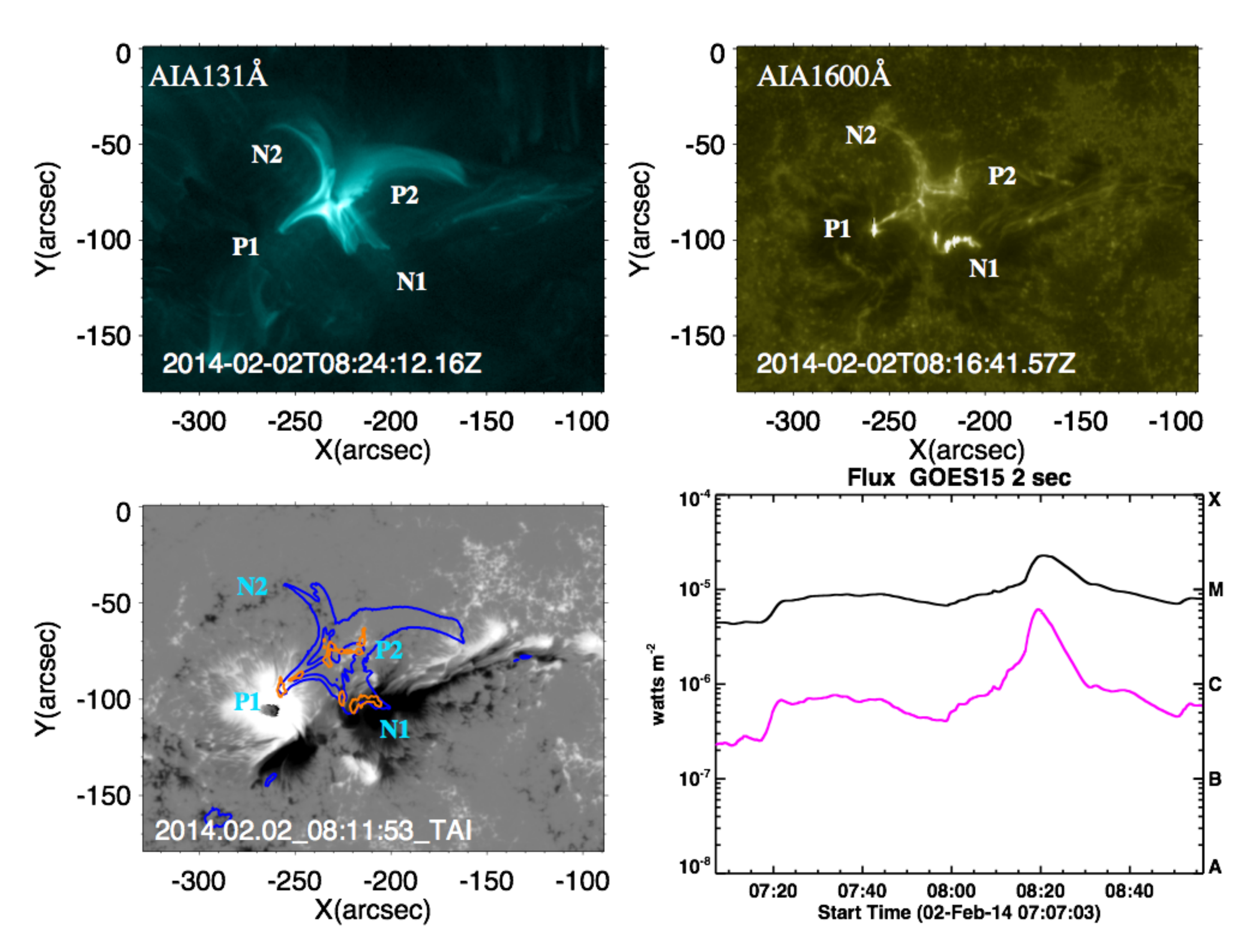}
 \caption{Upper left: An image of AIA 131 \AA \ in the main phase of flare 1, which was at peak at 08:20 on February 2, 2014. Upper right: AIA 1600 \AA \ image observed during the initial phase. Bottom left: the photospheric magnetic field with the contours of AIA 131 \AA \ and 1600 \AA \ (blue and orange, respectively). Bottom right: the time plots of the {\it GOES} X-ray flux (1-8 \AA \ and 0.5-4 \AA). North is up and west is to the right.}
\label{obs1}
\end{figure}

\begin{figure}
\centering
 \includegraphics[width=0.9\columnwidth,clip,bb=0 0 1034 793]{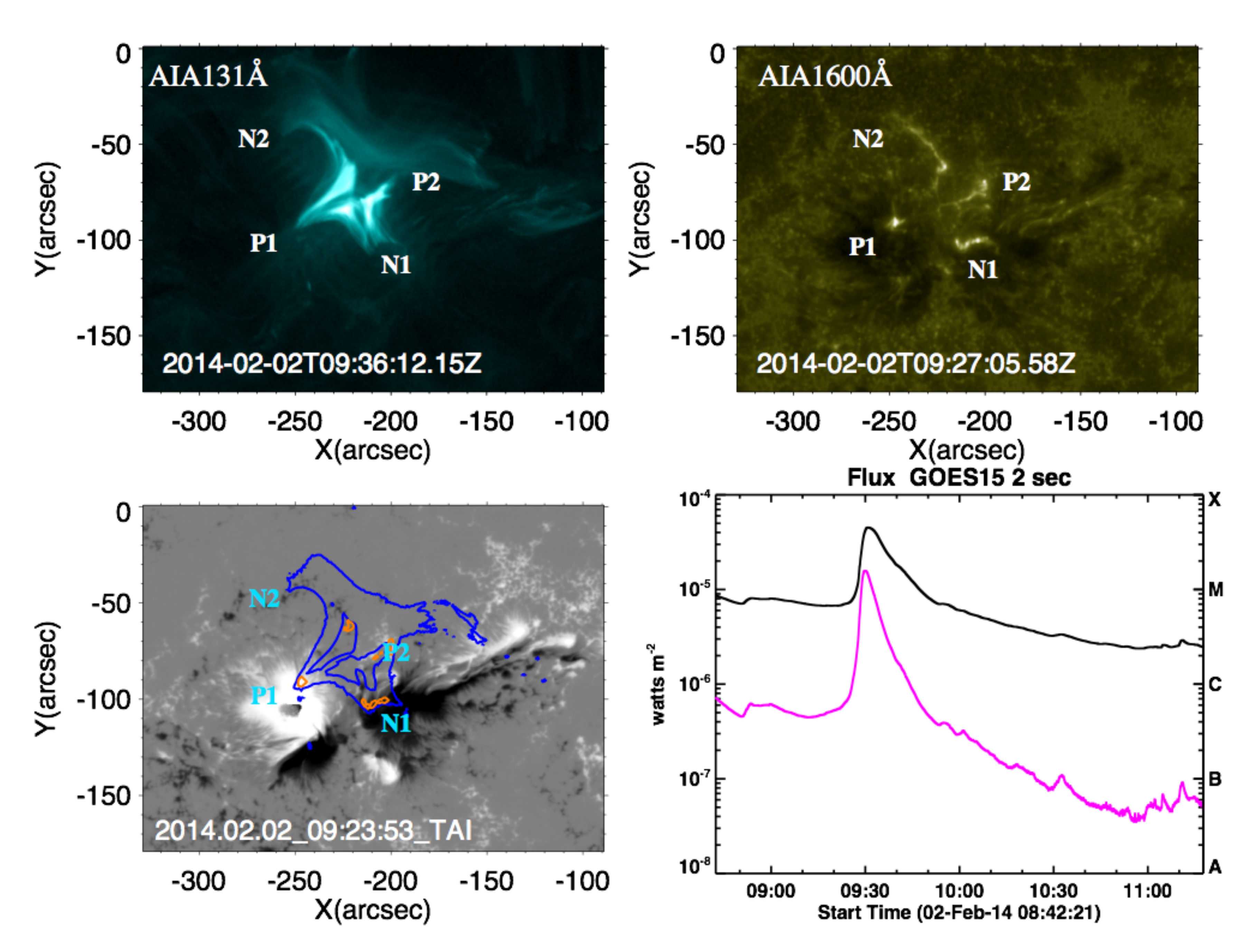}
 \caption{Similar to Figure \ref{obs1} for flare 2. The peak time was at 09:31 on February 2, 2014. }
\label{obs2}
\end{figure}

\begin{figure}
\centering
 \includegraphics[width=0.9\columnwidth,clip,bb=0 0 1034 793]{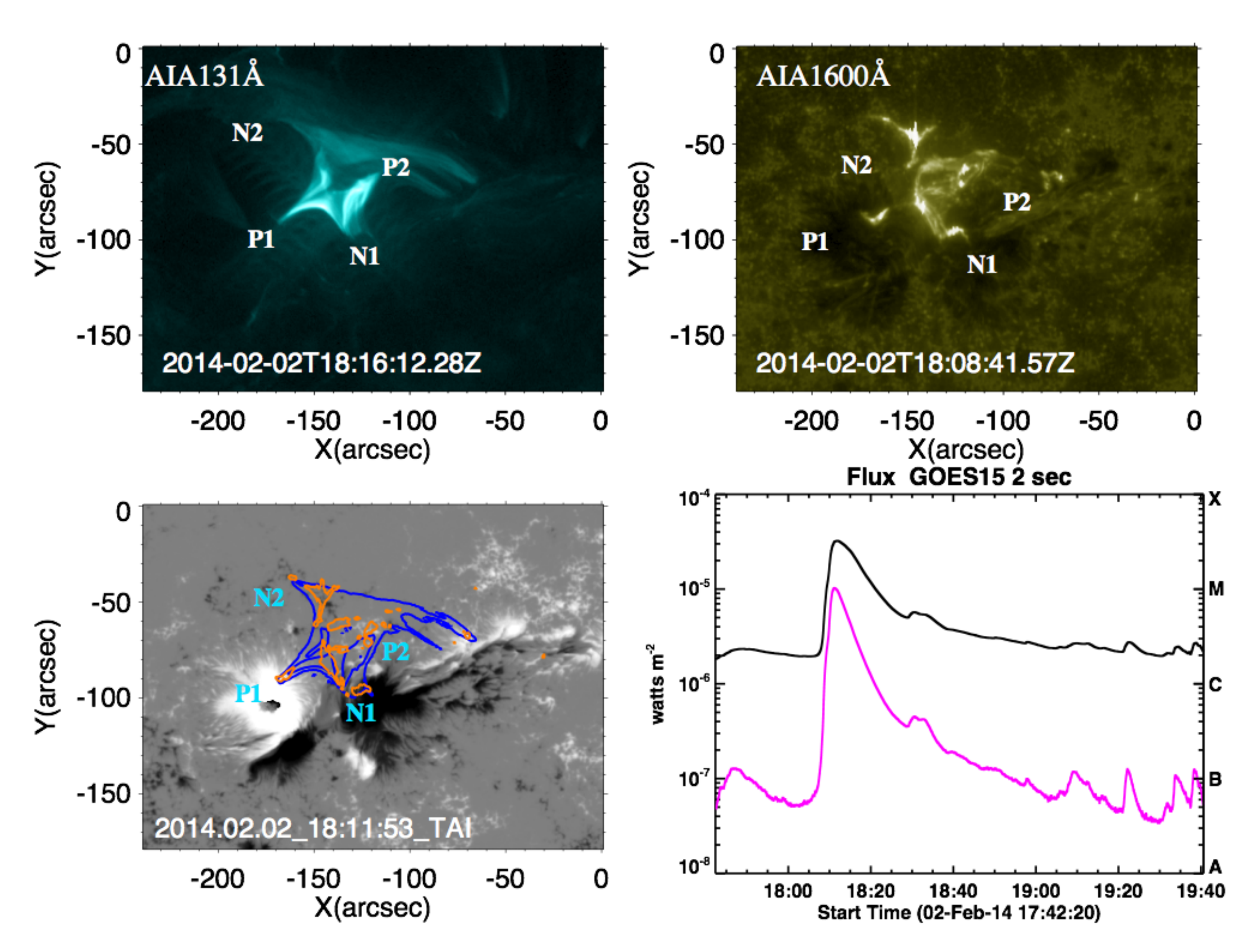}
 \caption{Similar to Figure \ref{obs1} for flare 3. The peak time was at 18:11 on February 2, 2014. }
\label{obs3}
\end{figure}

The EUV 131 \AA \ images obtained by {\it SDO}/AIA for the three flares are shown in the top left of Figures \ref{obs1}, \ref{obs2}, and \ref{obs3}, which provide the spatial distribution of hot coronal plasmas immediately after the main phase of each flare. The time plots of the {\it GOES} X-ray flux (bottom right) are shown on the bottom right of Figures \ref{obs1}, \ref{obs2}, and \ref{obs3}. 

When the flares occurred, flare ribbons appeared in the chromosphere. 
The AIA 1600 \AA \ images were used to identify the flare ribbons.
The upper right panel of Figures \ref{obs1}, \ref{obs2}, and \ref{obs3} show an AIA 1600 \AA \  image in the initial phase of each flare.
When flare 1 occurred, four flare ribbons appeared in response to the appearance of hot coronal plasmas. 
In Figure \ref{obs1}, we can identify the localized kernels of the flare ribbons located at two sunspots (P1, N1) and bright features located in the plage region (P2, N2). After the main phase, the P1-N2 and P2-N1 loops appeared with the pre-existing P1-N1 loops for all the three events in AIA 131 \AA. The significant changes in the connectivity of EUV bright structures near the time of the main phase suggest the magnetic reconnection in the coronal magnetic structures formed above the quadrupole magnetic polarity distribution. 
Figures \ref{obs2} and \ref{obs3} show flare ribbons similar to those of flare 1 in 1600 \AA , which indicates that they are characterized as homologous flares. Both flare 2 and flare 3 show four flare ribbons identified in the main phase.

\subsection{Force-free field modeling of NOAA 11967}
\begin{figure*}
 \centering
 \includegraphics[width=0.8\columnwidth,clip,bb=0 0 622 978]{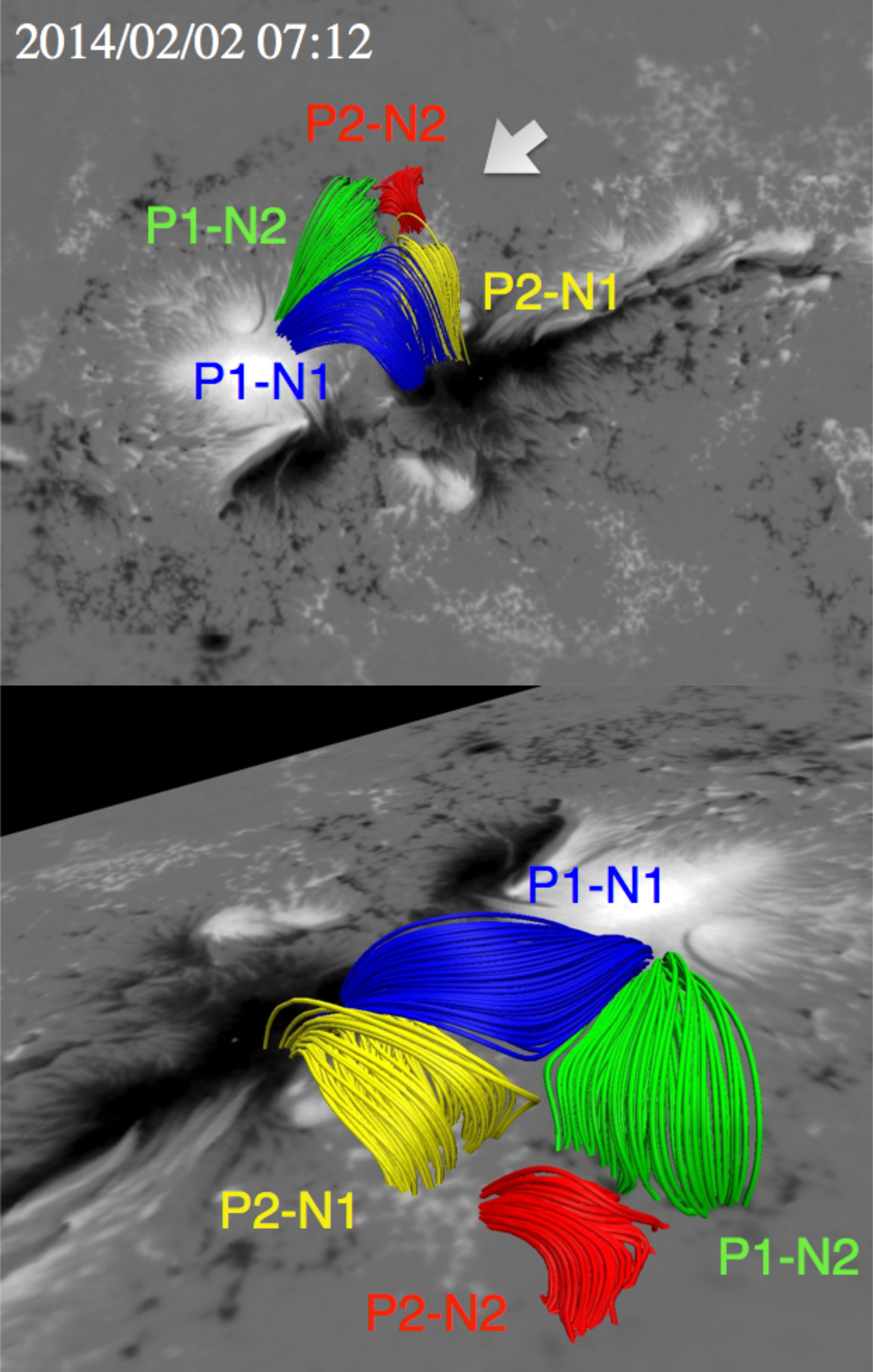}
 \caption{Result of the force-free field modeling based on SP data obtained between 07:12 and 08:07 UT on February 2, 2014. The colored lines show the magnetic field lines of P1-N1, P1-N2, P2-N2, and P2-N1. The gray arrow in the upper panel indicates the direction of view in the bottom panel. }
 \label{ff_result}
\end{figure*}

We performed NLFFF modeling using the method described in the previous section. First, we examined the consistency of the NLFFF modeling by comparing the footpoint connectivity of the field lines with the location of the flare ribbons observed by AIA 1600\AA. The force-free field modeling revealed four sets of connectivities in the magnetic field lines that originated in the four flare ribbons. One of the results of NLFFF modeling is shown in Figure \ref{ff_result}; it is  based on the SP data obtained between 07:12 and 08:07 UT on February 2, 2014, approximately one hour before the occurrence of flare 1.  The background map shows the vertical component of the magnetic field. The field lines are drawn from randomly selected points in each region. The blue, green, red, and yellow lines describes the magnetic field lines for the P1-N1, P1-N2, P2-N2, and P2-N1, respectively.  The gray arrow in the upper panel shows the direction of view in the bottom panel. 

\begin{figure}
\centering
 \includegraphics[width=0.9\columnwidth,clip,bb=0 0 590 817]{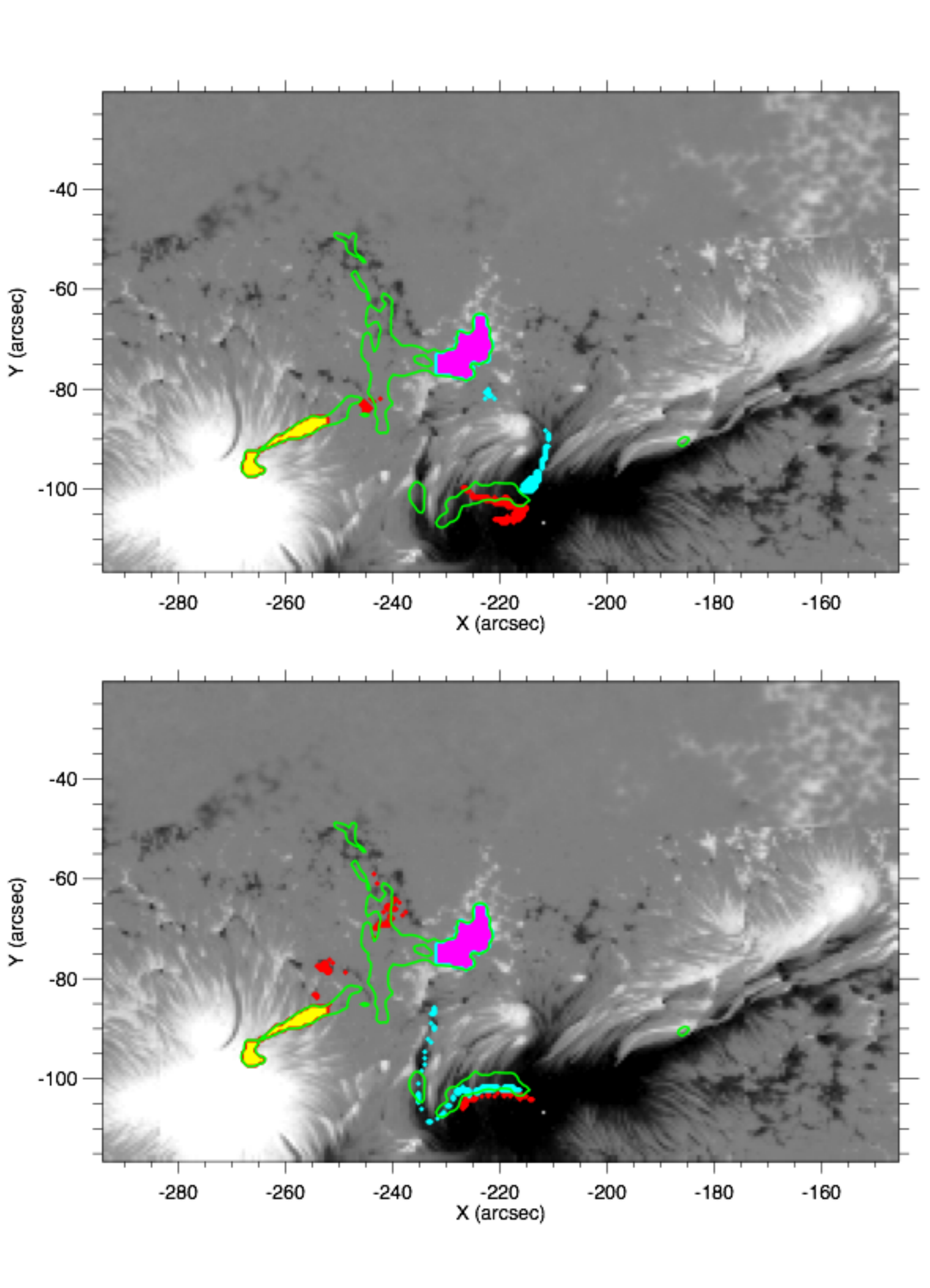}
 \caption{Comparison between the flare ribbons observed by AIA 1600 \AA \ (green contours) and the foorpoint pair of the field lines. The light blue area shows the footpoints of the magnetic field lines connection from the magenta region. Similarly, the red area shows the footpoints connected from the yellow region. Top panel: result of potential field. Bottom panel: result of NLFFF modeling.}
\label{check}
\end{figure}

Because flare ribbons result from transient heating at the chromospheric footpoints of the magnetic field lines associated with magnetic reconnection, the field lines from one flare ribbon could connect to another flare ribbon. By comparing the two locations, we can confirm the reliability of  the NLFFF modeling \citep{2011ApJ...738..161I}. In Figure \ref{check}, the flare ribbons of flare 1 observed by AIA 1600 are shown by the green contours on the vertical magnetic field map. The light blue filled area shows the footpoint of the magnetic field lines connecting from the magenta region. Similarly, the red filled area shows the footpoint  of the magnetic field lines connecting from the yellow region. The top panel shows a result of potential field modeling, whereas the bottom panel displays that of NLFFF modeling. Clearly, in potential field modeling, the field lines extending from the flare ribbons (magenta, yellow) do not connect to the flare ribbon located in the negative sunspot (N1). On the other hand, in NLFFF modeling,  the connectivity of field lines seems more consistent with the location and shape of the observed flare ribbon in the N1 region, although the red footpoints in the N1 region are slightly deviated from the location of the flare ribbon in the N1 region. This result suggests that the NLFFF modeling performed in this study can reproduce the connectivity of the field lines very well.

\subsection{Formation of non-potential magnetic fields}
\begin{figure*}
\centering
\includegraphics[width=0.9\columnwidth,clip,bb=0 0 1024 795]{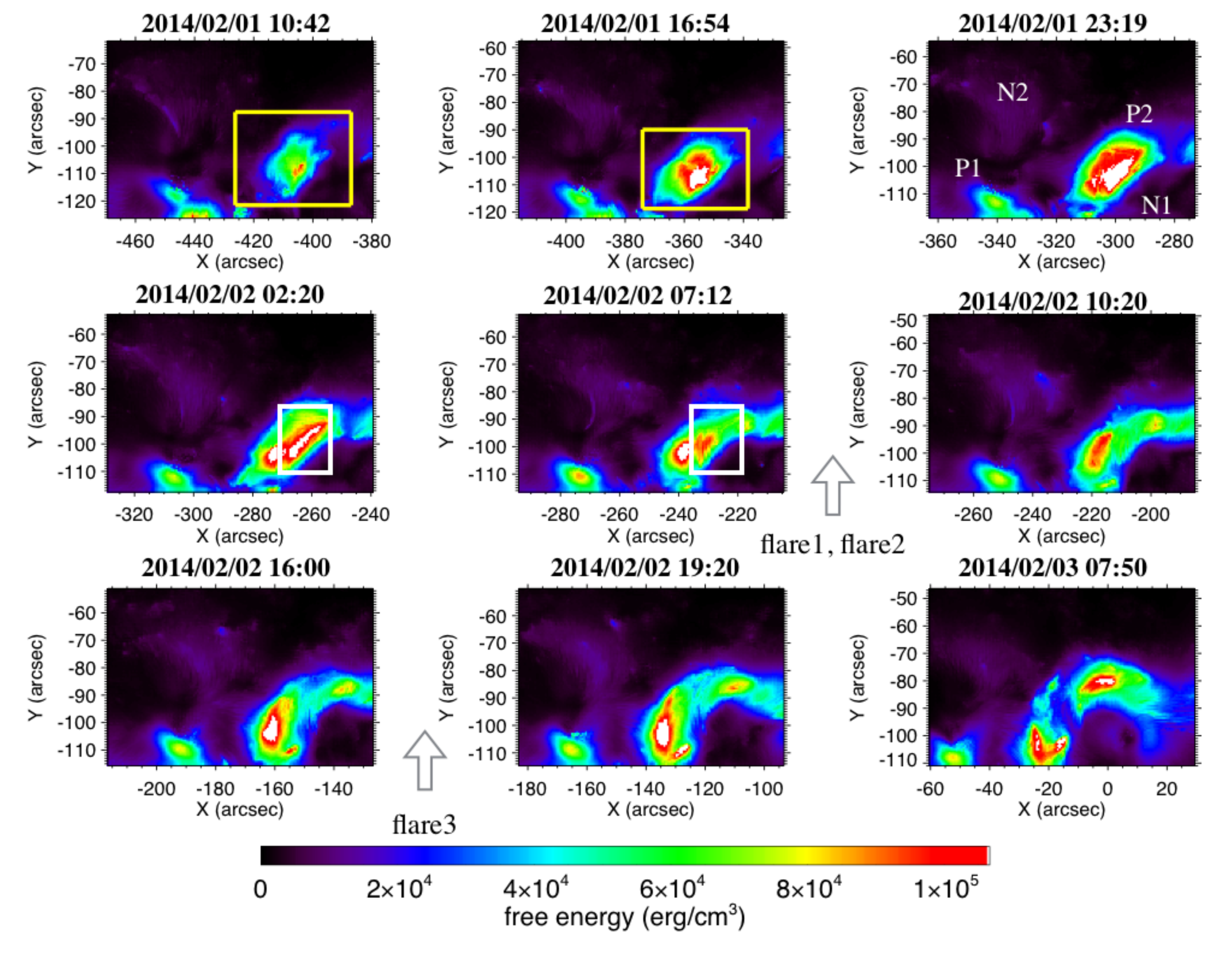}
\caption{Temporal evolution of the local free energy distribution averaged along the z-axis at each position. The FOV is shown by the red rectangle in Figure \ref{boundary}. The free energy, which was integrated in the yellow rectangle at 16:54 UT on February 1, 2014, was $4.7 \times 10^{31}$ erg. }
\label{fen_evo}
\end{figure*}

We analyzed the temporal evolution of the NLFFF based on the time series of the SP map. The energy for producing solar flares, i.e., the free energy, can be calculated from the results of the NLFFF modeling and is defined as
\begin{equation}
E_{\rm free}=\int \frac{B_{\rm nlfff}^2-B_{\rm potential}^2}{8\pi}dV .
\end{equation}
Figure \ref{fen_evo} shows the temporal evolution of the free energy density distribution, which is averaged along the vertical direction (z-axis) at each position. The FOV shown in Figure \ref{fen_evo} is indicated by the red rectangle in Figure \ref{boundary}. In the region between N1 and P2, which is enclosed in the yellow rectangle, the free energy was increased between 10:42UT and 16:54UT on February 1, 2014. The free energy integrated in the yellow rectangle at 16:54UT on February 1, 2014 was $4.7 \times 10^{31}$ erg, which is sufficient to produce an M-class flare. Subsequently, the free energy exhibited only a slight increase ($\sim$50$\%$ at most), although we cannot confirm whether these changes were more significant than the error bar.   Possible sources of errors include the error from the observations, the inversion of the spectropolarimetric data, the spatial resolution of the boundary data, and the NLFFF calculation method. Although the evaluation of such errors is important, it is difficult to consider all possible errors. This paper does not provide the uncertainty in the free energy.  Because the uncertainty should be evaluated, the evaluation of the errors will be reported in our subsequent paper. A decrease was also observed between 02:20UT and 07:12UT on February 2, 2014, as shown by the white rectangle in Figure \ref{fen_evo}. In summary, the present data show that the largest part of the free energy required for the M-class flares was stored more than 10 hours before the occurrence of the first M-class flare. The stored free energy was localized around the P2-N1 region and it was not distributed in the entire quadrupole field structure.

\begin{figure*}
\centering
\includegraphics[width=0.9\columnwidth,clip,bb=0 0 697 332]{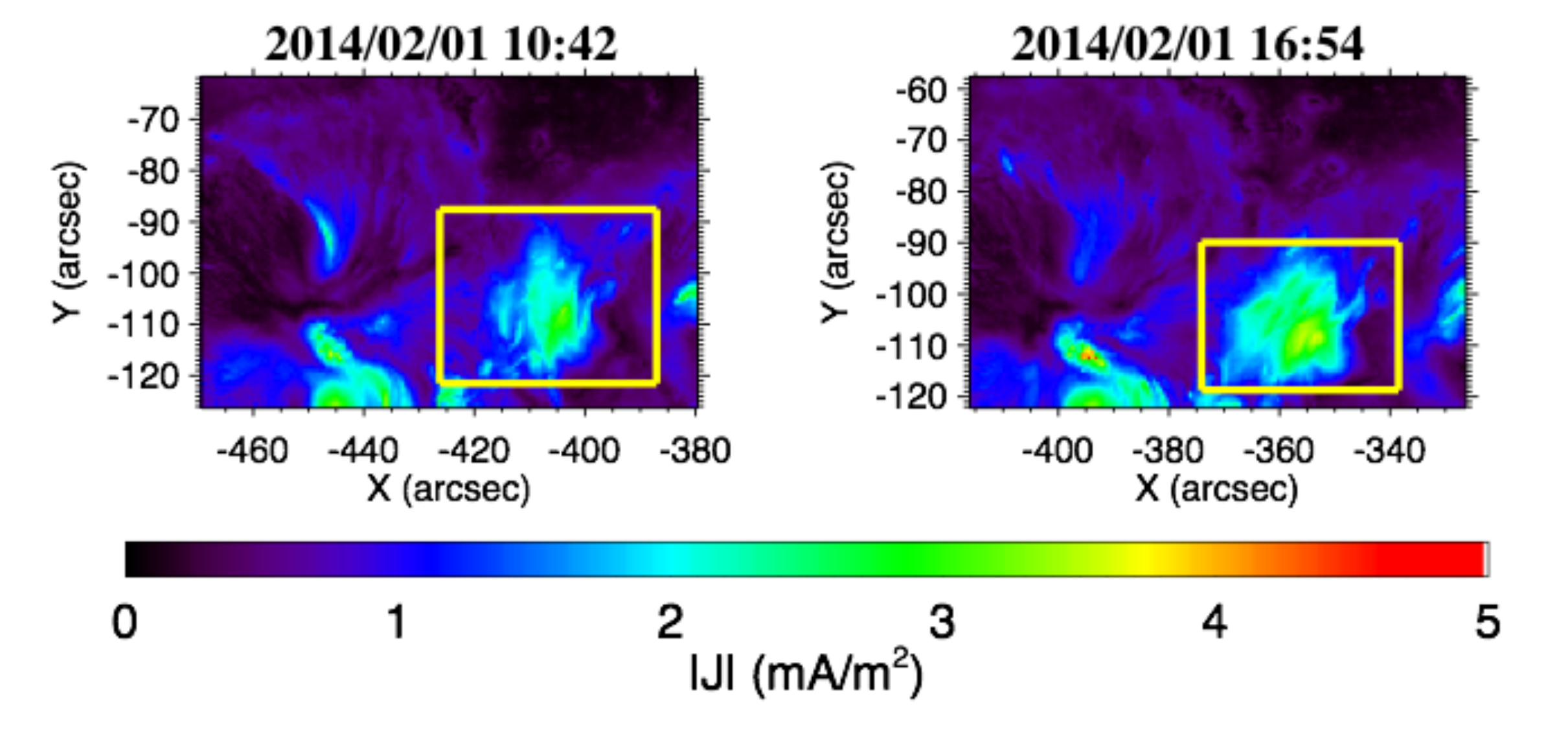}
\caption{Maps of the absolute value of the electric current distribution averaged along the z-axis at each position. The FOV is the same as that of Figure \ref{fen_evo}. }
\label{current_evo}
\end{figure*}

 Figure \ref{current_evo} shows maps of the absolute values of the electric current distribution at 10:42UT and 16:54UT on February 1, 2014, which were averaged along the z-axis at each position. The FOV is the same as that of Figure \ref{fen_evo}. The free energy results from the magnetic fields generated by the electric currents. The electric currents at a given point can create free energy everywhere and it is straightforward to interpret the maps of electric current as the difference from the potential configuration. However, the electric current in Figure \ref{current_evo} was also concentrated in the area inside the yellow rectangle simultaneously with the free energy in Figure \ref{fen_evo}. Therefore, in this case, we can deduce that the free energy density maps of Figure \ref{fen_evo} also contain the information of the distribution the of non-potential field configuration.

\begin{figure}
\centering
 \includegraphics[width=0.8\columnwidth,clip,bb= 0 0 1024 953]{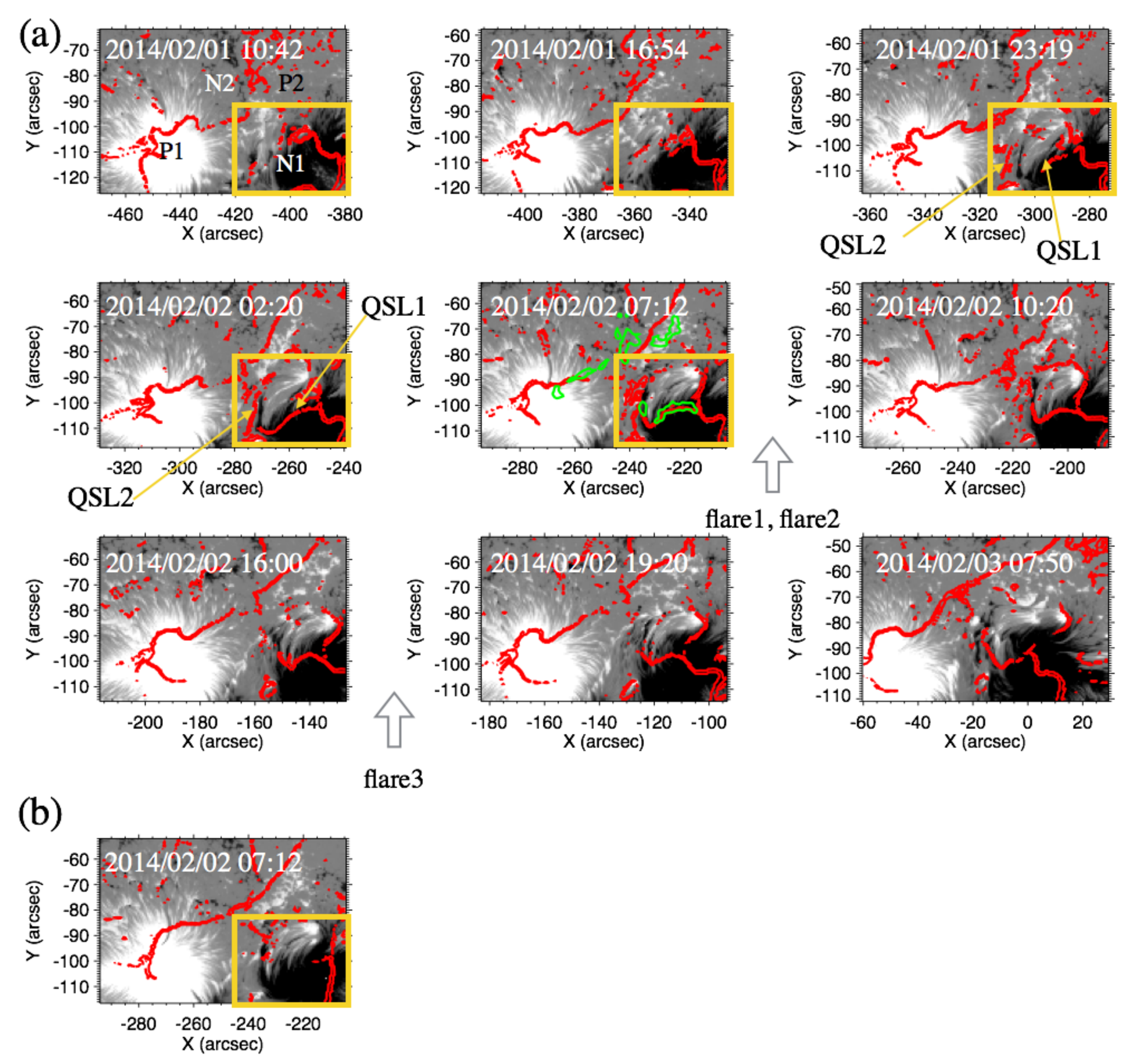}
 \caption{\small{(a) Temporal evolution of QSLs on the vertical magnetic field map. QSLs in the lower chromosphere ($\sim$ 400km above the formation height of Fe I line) are indicated by red contours ($\log Q>1$). The green contours show the  flare ribbons observed by AIA 1600\AA. The yellow rectangles show the region where the clear formation of QSLs can be seen. (b) QSLs distribution calculated using the potential field at 07:12 UT on February 2, 2014.}}
\label{qsl_evolution}
\end{figure}

Figure \ref{qsl_evolution}a shows the temporal evolution of the QSLs on a photospheric vertical magnetic field. The FOV is the same as that of Figure \ref{fen_evo}. The red contours indicate the location of  the QSLs ($\log Q>1$) in the lower chromosphere ($\sim$ 400km above the formation height of the Fe I line). To compare the location between the QSLs and flare ribbons, we have plotted the QSLs in the lower atmosphere. The QSLs in the N1 region that are enclosed by the yellow square show clear temporal changes. New QSLs (QSL1 and QSL2, indicated by the yellow arrows in Figure \ref{qsl_evolution}a) appeared by 23:19 UT on February 1, 2014, gradually became longer, and were eventually connected to each other by 07:20 UT, one hour before the occurrence of flare 1. The flare ribbons observed by AIA 1600\AA , shown by green contours in the 07:12 UT frame, are located  along QSL1 and QSL2. Note that the good correspondence between QSLs and flare ribbons has been reported in previous studies \citep{2015ApJ...810...96S,2016ApJ...818..168I}. After flare 1 and 2, the QSLs in the N1 region do not show significant temporal evolution. Even after flare 3, the QSLs remained in the N1 region.  For comparison, Figure \ref{qsl_evolution}b shows the location of the QSLs at 07:12  UT on 2 February 2014, which was derived from the potential field calculation. The QSLs in N1 can not be observed in the potential field.  

\begin{figure}
\centering
 \includegraphics[width=0.8\columnwidth,clip,bb= 0 0 854 568]{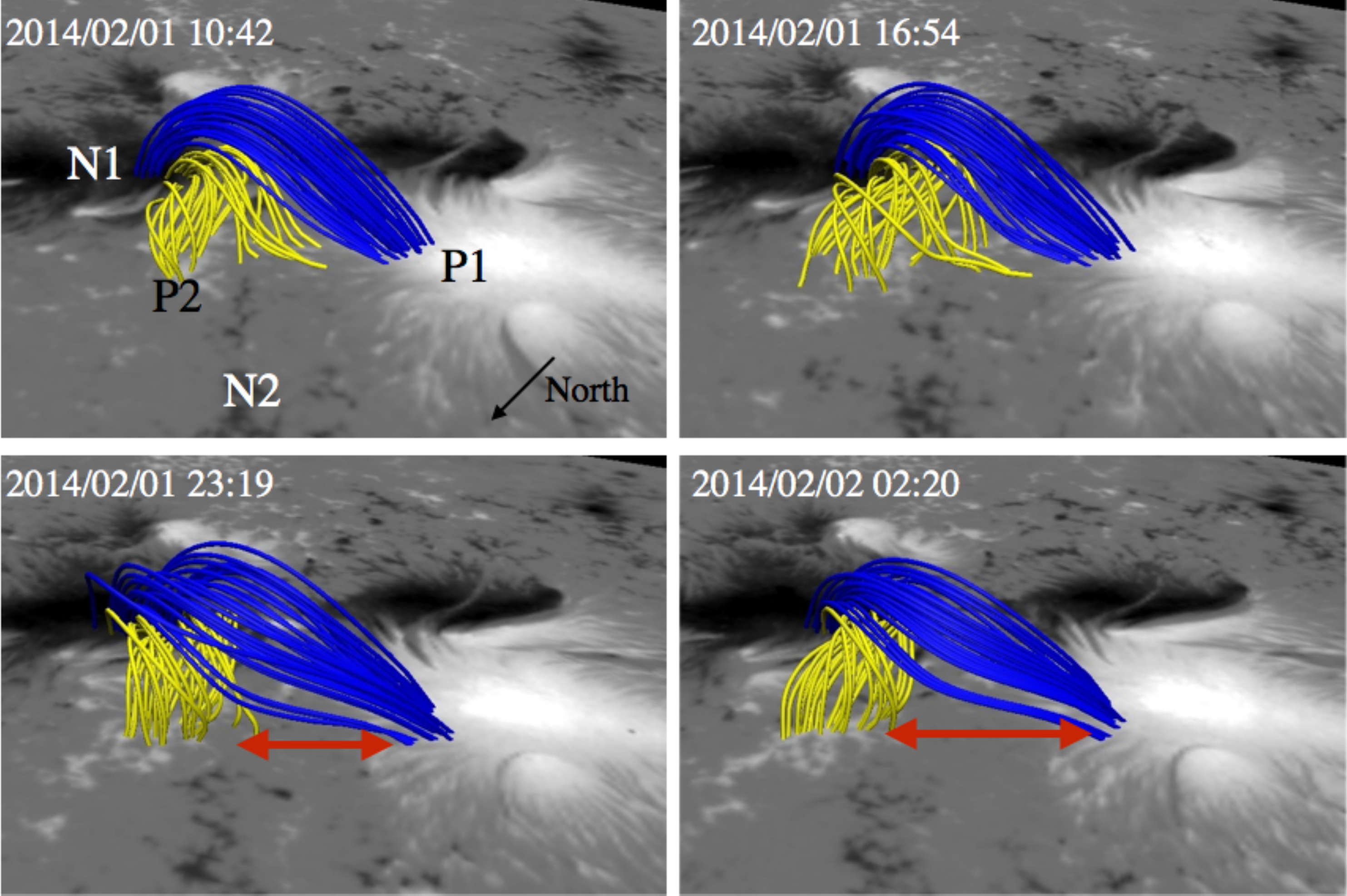}
 \caption{\small{Temporal evolution of the magnetic field lines P1-N1 and P2-N1, derived from the NLFFF modeling. The black arrow shows the direction of the north.}}
\label{p1p2evo}
\end{figure}

Figure \ref{p1p2evo} shows the temporal evolution of magnetic field lines P1-N1 (blue lines) and P2-N1 (yellow lines) near the formation time of  the new QSLs in N1.  At 10:42 UT and 16:54 UT on February 1,  2014, the positive footpoints of the P1-N1 and P2-N2 lines are continuously distributed and aligned to each other. At 23:19 UT on February 1 and 02:20 UT on February 2, 2014, the positive footpoints of the P2-N1 lines are clearly separated from those of the P1-N1 lines, creating a QSL between them.  

\begin{figure*}
\centering
\includegraphics[width=0.75\columnwidth,clip,bb=0 0 768 772]{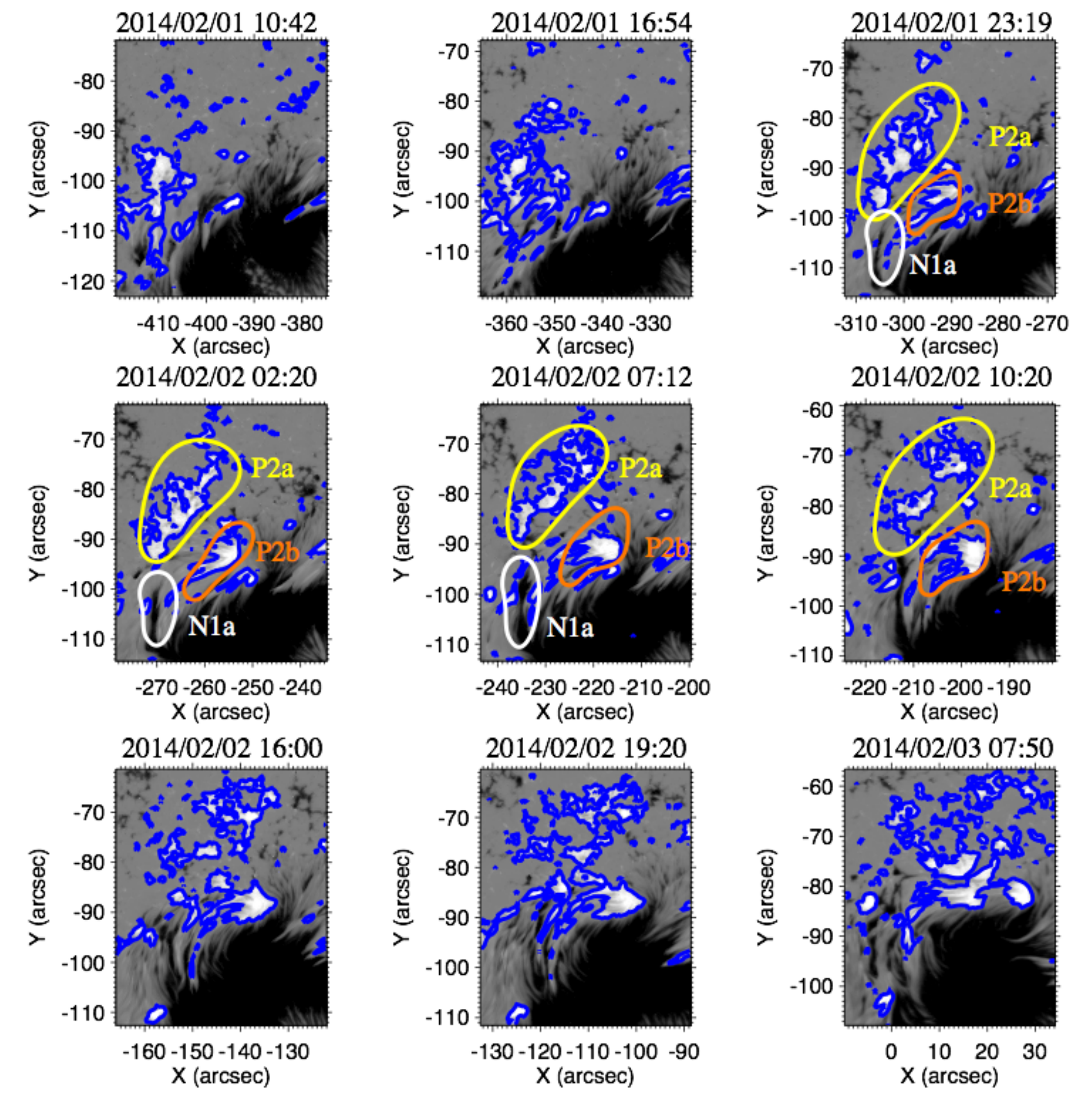}
\caption{Temporal evolution of the photospheric vertical magnetic field around the P2-N1 region. The blue contours indicate the location of the positive 300 G magnetic flux level. The FOV is shown by the blue rectangle in Figure \ref{boundary}.}
\label{photo_evo}
\end{figure*}

\begin{figure*}
\centering
\includegraphics[width=0.85\columnwidth,clip,bb=0 0 1700 850]{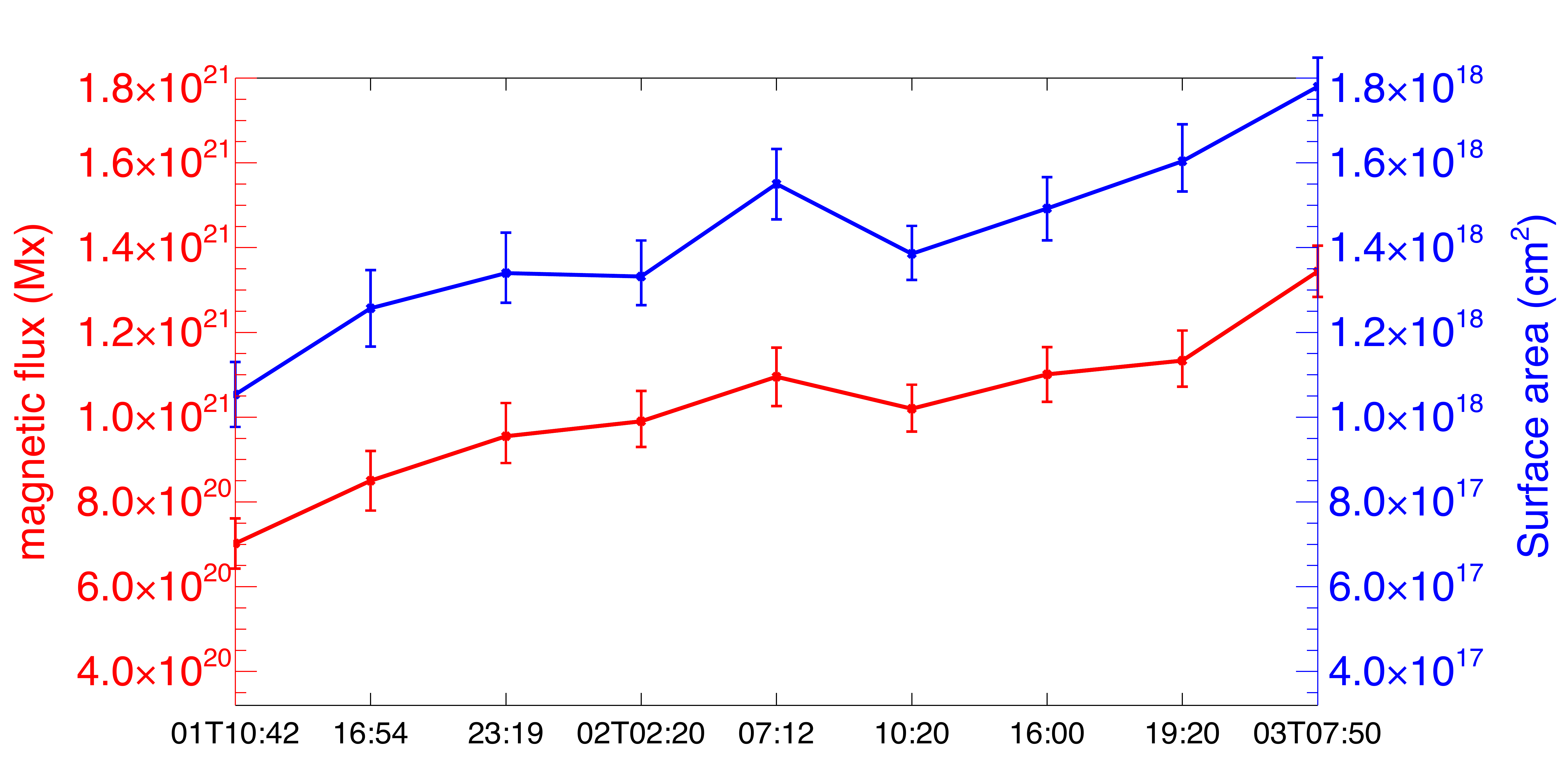}
\caption{Temporal evolution of the photospheric vertical magnetic flux (red solid line) and  surface area (blue solid lines) of  the positive magnetic magnetic field, which was measured with the blue contours shown in Figure \ref{photo_evo}. }
\label{photoflux_evo}
\end{figure*}

Figure \ref{photo_evo} shows the temporal evolution of the photospheric magnetic field distribution around the P2-N1 region. The FOV is indicated by the blue rectangle in Figure \ref{boundary}. The blue contours represent the positive polarity patches with the magnetic flux density of 300 G. Figure \ref{photoflux_evo} shows the temporal evolution of the vertical positive magnetic flux (red solid line) and  the surface area (blue solid line) measured with the blue contours of Figure \ref{photo_evo}. The vertical magnetic field flux increases by a factor of 2 and the surface area shows a similar increase. This increase indicates an emergence of magnetic flux in the P2-N1 region. 

\begin{figure*}
\centering
\includegraphics[width=0.9\columnwidth,clip,bb=0 0 608 353]{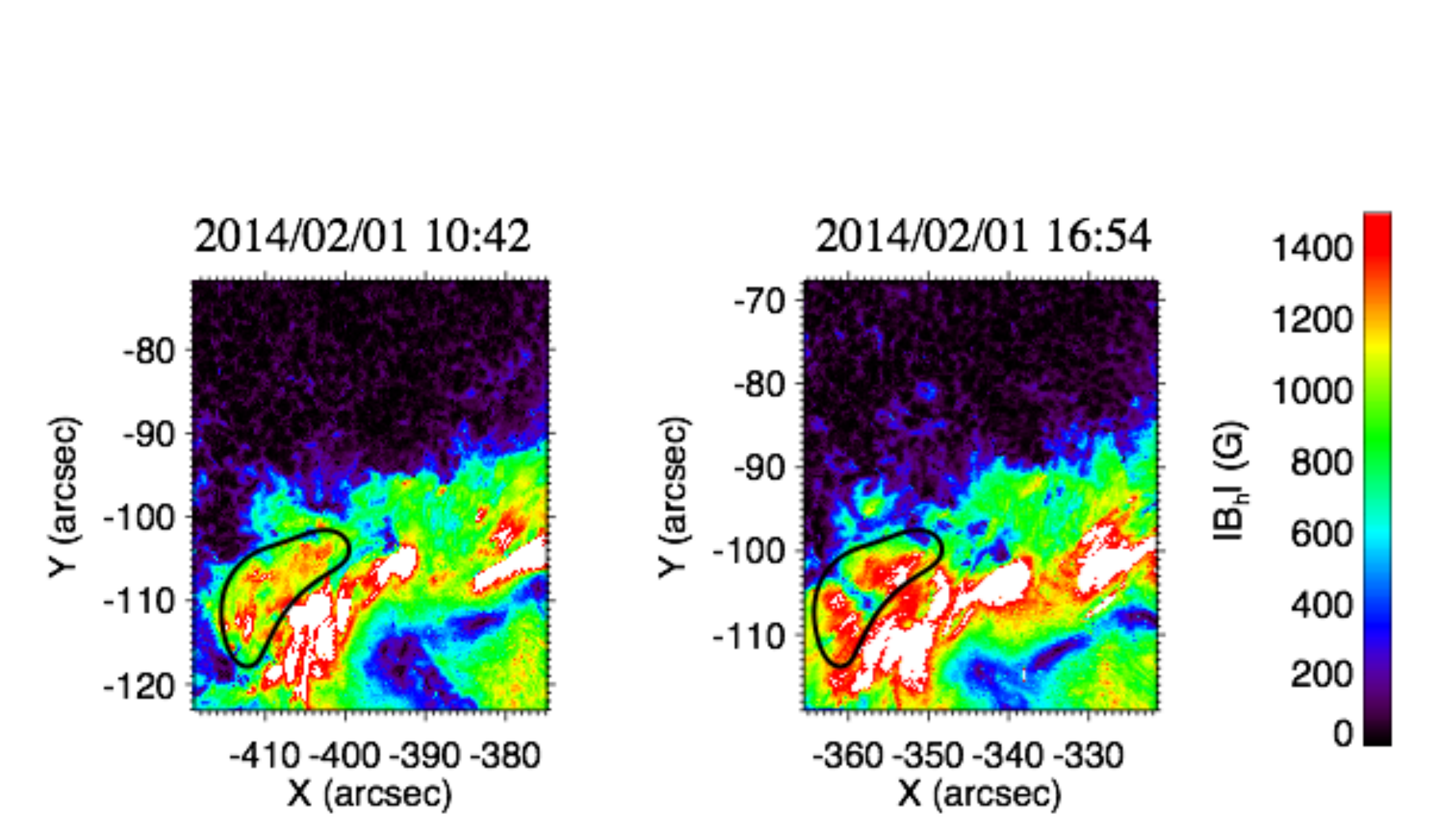}
\caption{Absolute value of the  horizontal component of the magnetic flux density. The increase of the  horizontal component can be seen in the closed black curve. The FOV is shown by the blue rectangle in Figure \ref{boundary}. }
\label{trans_evo}
\end{figure*}

 Although the free energy clearly increased between 10:42 UT and 16:54 UT on February 1, 2014,  the vertical magnetic fluxes in Figures \ref{photo_evo} and \ref{photoflux_evo} show a slight increase during that time. To discuss the origin of free energy, we present Figure \ref{trans_evo}, which shows the absolute value of the horizontal component of the magnetic flux density ($=\sqrt{(B_x^2+B_y^2)}$) in the P2-N1 region in the photosphere at 10:42 UT and 16:54 UT on February 1, 2014. The FOV is indicated by the blue rectangle in Figure \ref{boundary}, which is the same as that of Figure \ref{photo_evo}. An increase of the horizontal component can be seen in the black closed curve, which corresponds to the location where free energy increased.

\begin{figure*}
\centering
\includegraphics[width=0.9\columnwidth,clip,bb= 0 0 880 699]{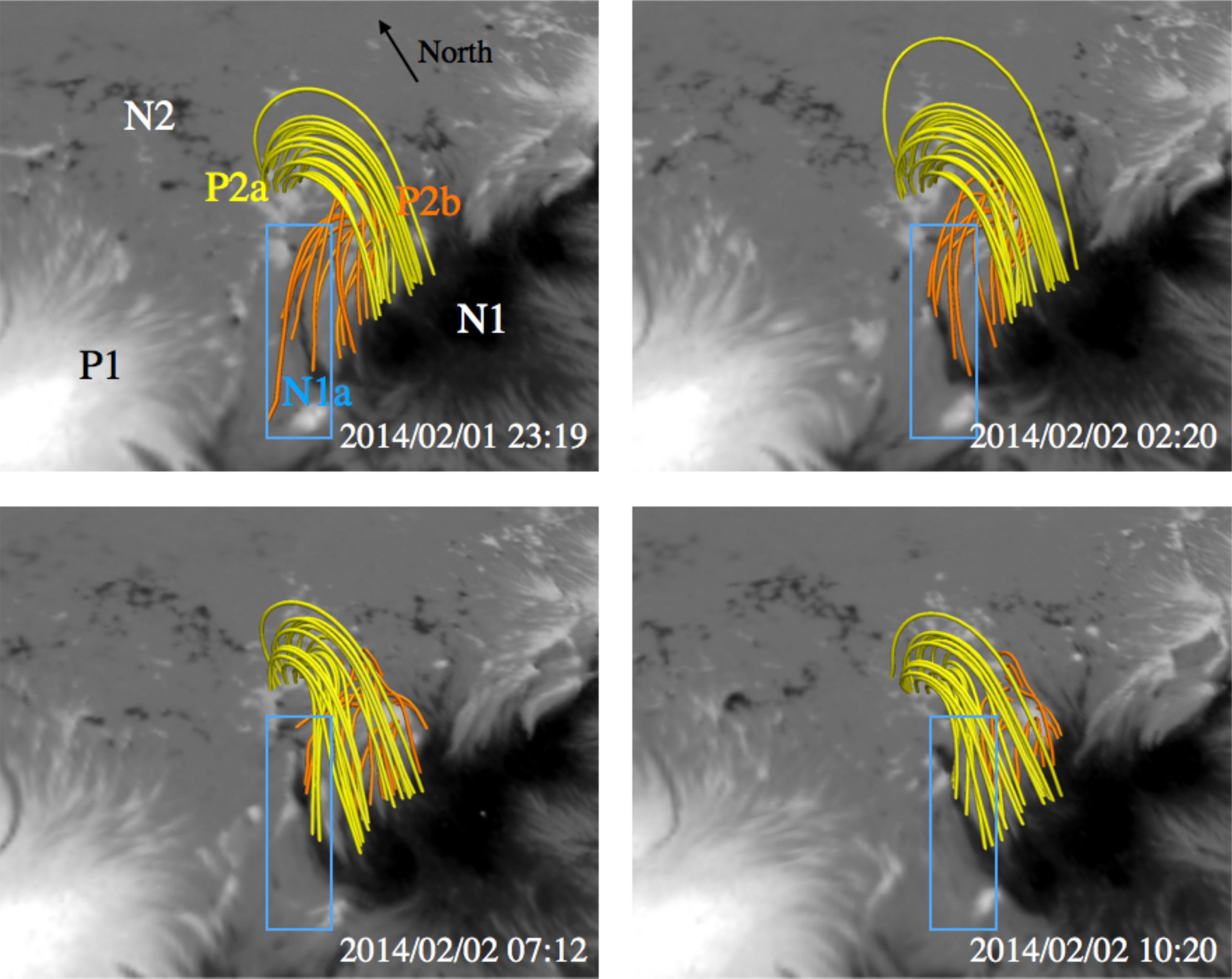}
\caption{Temporal evolution of the magnetic field lines derived from NLFFF modeling in the area containing N1a, P2a, and P2b. The magnetic field lines are plotted on the vertical magnetic field map, and the P2a-N1 and P2b-N1 field lines are shown in yellow and orange, respectively. The black arrow shows the direction of the north.}
\label{p2_evo}
\end{figure*}

The P2 region contains two groups of positive flux, P2a and P2b, which are shown by the yellow and orange circles in Figure \ref{photo_evo}.  In the N1 region, an elongated negative polarity field (N1a),  enclosed by the white circle, developed between 23:19 UT on February 1, 2014  and 07:12 UT on February 2, 2014. The N1a patch is a negative-polarity counterpart of the emerging flux with the positive-polarity P2b patch. Figure \ref{p2_evo} shows the temporal evolution of the magnetic field lines derived from NLFFF modeling in the area containing N1a, P2a, and P2b. The magnetic field lines are plotted on the vertical magnetic field map, and the P2a-N1 and P2b-N1 field lines are drawn in yellow and orange, respectively. The P2b lines were initially connected to N1a in the period between 23:19 UT on February 1, 2014 and 02:20 UT on February 2, 2014. However, the field lines connecting from the N1a region changed from P2b (orange lines) to P2a (yellow lines) at 07:12 UT on February 2, 2014. This transition may require the magnetic reconnection between P2a-N1 lines and P2b-N1 lines. 

\section{Discussion}

We performed temporal analysis of a 3D magnetic field using NLFFF modeling and focusing on the formation of the non-potential magnetic field. As shown in Figure \ref{fen_evo}, the free energy was locally supplied to the region between N1 and P2, rather than to the entire quadrupole structure, more than 10 hours before the first M-class flare.  We suspect that the flux emergence  was the source of free energy. According to Figure \ref{photo_evo}, we can not identify vertical flux in P2b and N1a at 16:54 UT on February 1, 2014, which clearly indicates a flux emergence after that time. This implies that the magnetic field in P2b and N1a was inclined at that time because the magnetic field in the region was in the initial phase of the flux emergence. This interpretation is supported by the result shown in Figure \ref{trans_evo}. An increase in the horizontal component can be seen in the region, in which the free energy increased between 10:42 UT and 16:54 UT on February 1, 2014, which is consistent to the interpretation of the initial phase of flux emergence.   Note that the magnitude of the horizontal component seems to increase in the other locations, in which no increase in the free energy could be seen. This suggests that the non-potential flux emergence resulted in the large free energy supply between the N1 and P2 regions, while the increase in the horizontal component in the other regions was due to the emergence of a potential-like magnetic field and did not contribute to the storage of free energy. On the other hand, the QSLs exhibit a different temporal evolution from that of the free energy. As reported by \cite{2016arXiv160902713L}, our QSL analysis also shows the existence of X-shaped QSLs at the center of the quadrupolar structure in NOAA11967 before the first M-class flare.  We revealed that the part of the X-shaped QSLs, i.e., QSLs in the N1 region, started to form from 9 hours before the first M-class flare, whose timing was different from that of the free energy storage. The QSLs derived from the potential field did not appear in the N1 region. These results indicate that the QSLs at the N1 region were products of non-potential magnetic field structures, which were necessary to produce the flares and that the high free energy alone can not produce the flares.

Let us discuss the mechanisms of the formation of the QSLs in the N1 region. There are two candidates for the formation of the new QSLs; transverse photospheric motion and emerging flux. We identified observational signatures supporting both of them. We consider that the formation of the east-west QSL (QSL1)  and the north-south QSL (QSL2) in the N1 region were due to the transverse photospheric motion of the pre-emerged flux and the emergence of a new flux, respectively. The footpoint transverse motion of the P2-N1 line is shown in Figure \ref{p1p2evo}. Initially, the footpoints of P1-N1 and P2-N1 were aligned and there was no clear spatial separation between P1 and P2. The P2 region moved toward the north-west direction (outward from the P1 region), as also seen in Figure \ref{photo_evo}. The distance between P1 and P2 gradually increased, resulting in the formation of QSL1. The flux emergence is evident in Figures \ref{photo_evo} and \ref{photoflux_evo}. The surface area and the flux increased with time. The appearance of the N1a flux was clearly recognized at 23:19 UT on February 1, 2014. According to the NLFFF result, the counterpart of the N1a region was initially P2b, suggesting that the N1a-P2b pair was an emerging flux. The appearance of the N1a region created QSL2 between the field lines connecting to P1 and the field lines to P2b.

We suspect that the occurrence of the first M-class flare was related to the formation of QSL1. Before the formation of QSL1, the magnetic field configuration was more similar to a tripole rather than a quadrupole because the distance between P1 and P2 was short. This means that the anti-parallel component between P1-N2 and P2-N1 was small, i.e., magnetic reconnection was difficult to occur. The transverse motion of P2 increased the anti-parallel component and enabled the magnetic reconnection.

We have shown that the magnetic free energy is stored inside the X-shaped quadrupole structure. It is locally stored around the N1 region, where two new QSLs were clearly developed at least one hour before the onset of the first flare (flare1). This suggests that in addition to the storage of free energy in the corona, the formation of QSLs in the free energy area may be also required to produce the flares. However, the QSLs were clearly observed in the N1 region even one day later, i.e., at 07:50 UT on February 3, 2014. Even in such a circumstances, no flares were produced in this quadrupole region on February 3, 2014 or later. This may indicate that another condition is necessary to produce the flare.  One such candidate may be the change of the connectivity  between P2a-N1 and P2b-N1a, as shown in Figure \ref{p2_evo}. This result suggests that magnetic reconnection occurred between P2a-N1 and P2b-N1a. The magnetic reconnection process converts the magnetic free energy to other types of energy. In other words, magnetic reconnection acts as the local relaxation of the magnetic energy. This magnetic energy relaxation may be observed as a slight decrease in the free energy, as shown in Figure \ref{fen_evo}. The free energy in the P2 region, which is enclosed by the white rectangle, seemed to decrease between 02:20 UT and 07:12 UT on February 2, 2014.  The occurrence of solar flares (or solar eruptions) has been described by a variety of storage-and-release models \citep{2014IAUS..300..184A}. Although the increase of free energy is necessary for the occurrence of solar flares, only the storage of energy can not explain all the trigger mechanisms of solar flares. For example, the tether-cutting model \citep{2001ApJ...552..833M} and the breakout model \citep{1999ApJ...510..485A} require magnetic reconnection as a precursor before solar flares, which contributes to decreasing magnetic tension and leads to the rise of flux rope (MHD instability) and solar flares. In our case, we propose the decrease in free energy before the first M-class flare, which suggests that magnetic reconnection in a localized region might  also contribute to triggering the flares. The local magnetic reconnection in the P2-N1 region caused a decrease in the magnetic pressure, which may behave as a disturbance to the force-free equilibrium in the entire quadrupolar system and lead to the occurrence of the flare. 

\section{Conclusions}
We investigated the temporal evolution of the 3D magnetic field structure in an X-shaped quadrupolar magnetic field region producing M-class solar flares. Firstly, we evaluated the accuracy of NLFFF modeling by comparing the location of the flare ribbons observed with AIA 1600 with the footpoint connectivity of the field lines. While the footpoints of the field lines from the potential field were not co-spatial with the flare ribbons, the NLFFF modeling results exhibited the good correspondence with the location of the flare ribbons. This demonstrates that the NLFFF modeling can reproduce the coronal magnetic field much better than potential field modeling. 

The formation of the non-potential field in the X-shaped  quadrupolar structure was investigated in this study. We performed NLFFF modeling by using a time series of vector magnetic field maps from {\it Hinode} SOT/SP and SDO/HMI. Our analysis revealed the following.
The free energy had already been stored in the localized region between the N1 and P2 regions more than 10 hours before the occurrence of the first M-class flare. Since one of the flare ribbons was co-spatial with the QSLs, the energy release process was related to the formation of the QSLs in the N1 region. We revealed that the QSLs in the N1 region started to form from 9 hours before the occurrence of the first M-class flare, and that they were finally connected to each other approximately one hour before the flare occurrence. The new QSL structure can not be seen in the potential field, indicating that the non-potential field created the new QSLs. We determined that both the transverse photospheric motion of the pre-emerged flux and the emergence of a new flux  were important for QSL formation. The transverse motion in the photosphere separated the two positive regions, P1 and P2. The evidence of the flux emergence was provided by the increase of the photospheric magnetic flux. This observation indicates that the occurrence of the flares requires the formation of QSLs in the non-potential field where the free magnetic energy is stored in advance.

{\bf Acknowledgements}
We thank the referee for constructive criticisms and valuable comments, allowing us to improve the manuscript significantly. {\it Hinode} is a Japanese mission developed and launched by ISAS/JAXA, collaboratoing with NAOJ as a domestic partner, NASA and STFC (UK) as international partners. It is operated by these agencies in cooperation with ESA and NSC (Norway). We gratefully acknowledge the {\it SDO}/HMI and {\it SDO}/AIA team for providing data. Our NLFFF calculations were performed on JAXA Supercomputer System generation 2 (JSS2). This work was supported by MEXT/JSPS KAKENHI Grant Numbers JP15H05814, JP15K21709 and JP15H05750.

\bibliographystyle{apj}


\clearpage



\clearpage




\end{document}